\documentstyle[aasms4,psfig,raulbib]{article}

\begin{document}

\title{\bf Proto--Galactic Starbursts at High Redshift}

\author{Raul Jimenez\altaffilmark{1}, Paolo Padoan\altaffilmark{2}, James
S. Dunlop\altaffilmark{1}, David V. Bowen\altaffilmark{3}, 
Mika Juvela\altaffilmark{4}, Francesca Matteucci\altaffilmark{5}}

\affil{$^1$Institute for Astronomy, University of
Edinburgh, Blackford Hill, Edinburgh EH9-3HJ, UK.}  
\affil{$^2$Harvard University Department of Astronomy, 60 Garden Street, 
Cambridge, MA 02138, USA.}
\affil{$^3$Princeton University Observatory, Princeton, NJ 08544, USA.}  
\affil{$^4$Helsinki University Observatory, T{\"a}htitorninm{\"a}ki,
P.O.  Box 14, SF-00014 University of Helsinki, Finland.}  
\affil{$^5$Department of Astronomy, University of Trieste, SISSA/ISAS, 
Via Beirut 2-4, 34014 Trieste, Italy.}

\authoremail{raul@roe.ac.uk}

\begin{abstract}
We have computed the evolving ultraviolet-millimeter spectral energy
distribution (SED) produced by proto-galactic starbursts at high redshift,
incorporating the chemical evolution of the interstellar medium in a
consistent manner. Dust extinction is calculated in a novel way, that is not
based on empirical calibrations of extinction curves, but rather on the
lifetime of molecular clouds which delays the emergence of each successive
generation of stars at ultraviolet wavelengths by typically 15 Myr. The
predicted rest-frame far-infrared to millimeter-wave emission includes the
calculation of molecular emission-line luminosities ($^{12}$CO and O$_{2}$
among other molecules) consistent with the evolving chemical abundances.  Here
we present details of this new model, along with the results of comparing its
predictions with several high-redshift observables, namely: the ultraviolet
SEDs of Lyman-limit galaxies, the high-redshift radiogalaxies 4C41.17 and
8C1435, the SCUBA sub-millimeter survey of the Hubble Deep Field (HDF), and
the SEDs of intermediate-redshift elliptical galaxies.  With our new reddening
method, we are able to fit the spectrum of the Lyman--limit galaxy 1512-cB58,
and we find an extinction of about 1.9 magnitudes at 1600 {\AA}. This
extinction apply to starbursts with spectral slope $\alpha$ in the range $0
\la \alpha \la 1.5$. The model also predicts that most Lyman--limit galaxies
should have a value of $\alpha$ inside that range, as it is observed.  The 850
$\mu$m flux density of a typical Lyman-limit galaxy is expected to be only
$\simeq 0.5$ mJy, and therefore the optical counterparts of the most luminous
sub-mm sources in the HDF (or any other currently feasible sub-mm survey) are
unlikely to be Lyman-break galaxies.  The passive evolution of our starburst
model is also compared with Keck observations of the reddest known elliptical
at $z \sim 1.5$, and with the SED of a typical nearby elliptical galaxy.  The
SED of the high redshift elliptical is nicely matched by the starburst model
with an age of 4 Gyr, and the SED of the nearby elliptical with an age of 13
Gyr.
\end{abstract}


\section{Introduction}
Optical surveys of normal and active galaxies have recently begun to provide
meaningful constraints on the star formation history of the Universe
\cite{M97,Dunlop_sfr98,PKSDAG98,S+98}. There is now substantial evidence that
the star formation rate grows \cite{M97,PKSDAG98} with increasing look-back
time out to a redshift of $z \simeq 2$, but calculations of the level of
star-formation activity at even higher redshift are still being
refined. Initial analyses of the optical image of the HDF indicated a clear
decline in star-formation density beyond $z \simeq 1.5$ \cite{M97}, but new
sub-mm observations of the HDF with SCUBA \cite{H+98} and new high-redshift
Lyman-limit galaxy samples \cite{S+98}, provide little evidence for any clear
decline between $z = 2$ and $z = 4$. Furthermore, it is possible that an
important population of high-redshift star-forming galaxies still remains
undetected by optical surveys (\pcite{AZ99}), due to being heavily embedded in
a significant amount of dust.

The best candidates for high-redshift dust-enshrouded starbursts are objects
in which star formation takes place before the gas is able to settle into a
disk. It can be argued that these may be the progenitors of present-day
elliptical galaxies and bulges, because the stellar populations of elliptical
galaxies have properties suggestive of a short (about 1 Gyr or less) formation
time-scale, involving a rapid build-up of metals and perhaps dust.  Even in
the hierarchical galaxy-formation scenario, sub-galactic size clumps are
expected to form stars well before these sub-units merge into larger
galaxy-size objects (e.g. \pcite{PJJ97}), and the star-formation in these
sub-units may be extremely vigorous. In a CDM cosmology small halos will
collapse at high redshift ($z>3$) and will have a high initial surface density
(e.g. \pcite{JPMH98}).  If the star-formation rate is well described at these
high redshifts by the Schmidt law \cite{K98}, a straightforward computation
shows that they will exhaust their gas in less than 1 Gyr \cite{HJ99}, thus
one expects these systems to behave like ``mini-starbursts'' that will quickly
form stars. These sub-units will subsequently merge to form the bulges of
disks or indeed elliptical galaxies.

In this work we do not therefore refer to a specific model for galaxy
morphological evolution, but rather study the properties of generic
star-formation bursts that occur at high redshift ($z>3$), and last less than
1 Gyr. Such bursts enrich the interstellar gas to solar metallicity in only
0.2 Gyr, and they might set up strong winds, able to expel most of the gas, in
about 0.3--0.8 Gyr.

The purpose of this paper is to present a new self--consistent model for the
ultraviolet--to--millimeter spectral energy distribution (SED) of
proto-galactic starbursts in which the far--infrared (FIR) dust emission
beyond 100 $\mu$m and molecular emission line luminosities are calculated
consistently with the radiation field and chemically evolving ISM produced by
the rapidly evolving stellar population.  These models build on new advances
in constructing synthetic stellar populations \cite{JPMH98,J+99} and new
three-dimensional simulations of molecular-cloud turbulence, that
realistically reproduce the properties of Galactic giant molecular clouds
(GMC) \cite{PJBN98} and allow an accurate modeling of the molecular lines,
using a non-LTE radiative transfer code.  The model is described in \S 2 to \S
5, and some general results are presented in \S 6. The results of the model
are compared with data from galaxies at high redshift in \S 7, and conclusions
are drawn in \S 8.

\section{Synthetic Stellar Population Models}

In order to compute the spectro-photometric evolution of starbursts we have
used the set of synthetic stellar population models developed by us
(\pcite{JPMH98}).  Our models are based on the extensive set of stellar
isochrones computed by \scite{JM96} and the set of stellar photospheric models
computed by \scite{Kurucz_92} and \scite{J+99}. The interior models were
computed using JMSTAR15 (\pcite{J+99}) which uses the latest OPAL95 radiative
opacities for temperatures larger than 8000 K, and Alexander's opacities
(private communication) for those below 8000 K. For the stellar photospheres
with temperatures below $8000$ K we have used a set of models computed with an
updated version of the MARCS code (U. J{\o}rgensen, private communication),
where we have included all the relevant molecules that contribute to the
opacity in the photosphere. For temperatures larger than $8000$ K we have used
the set of photospheric models by \scite{Kurucz_92}.  Stellar tracks were
computed self-consistently, i.e. the corresponding photospheric models were
used as boundary conditions for the interior models. This procedure has the
advantage that, since the stellar spectrum is known at any point on the
isochrone, the interior of the star is computed more accurately than if a grey
photosphere were used, and therefore we overcome the problem of using first a
set of interior models computed with boundary conditions defined by a grey
atmosphere and then a separate set of stellar atmospheres, either observed or
theoretical, that is assigned to the interior isochrone a posteriori. The
problem is most severe if observed spectra are used, because metallicity,
effective temperature and gravity are not accurately known and therefore their
position in the interior isochrone may be completely wrong (in some cases the
error is larger than 1000 K).  A more comprehensive discussion and a detailed
description of the code can be found in \scite{J+99}.

An important ingredient in our synthetic stellar population models is the
novel treatment of all post-main evolutionary stages that incorporates a
realistic distribution of mass loss. As a result, the horizontal branch is an
extended branch and not a red clump like in most stellar population
models. Also the evolution along the asymptotic giant branch is done in a way
such that the formation of carbon stars is properly predicted, together with
the termination of the asymptotic giant branch. In most synthetic stellar
population models a constant mass loss law is applied resulting in horizontal
branches that are simply red clumps. This leads to synthetic populations that
are too red by 0.1 to 0.2 in e.g. $B-R$ (\pcite{J+99}).

Using the above stellar input we compute synthetic stellar population models
that are consistent with their chemical evolution of the interstellar
medium. The first step is to build simple stellar populations (SSPs). A SSP is
a population of stars formed all at the same time and with homogeneous
metallicity. The procedure to build an SSP is the following:

\begin{enumerate}
\item A set of stellar tracks of different masses and of the metallicity
of the SSP is selected from our library.
\item The luminosity, effective temperature and gravity at
the age of the SSP is extracted from each track. Each set
of values characterizes a star in the SSP.
\item The corresponding self-consistent photospheric model is assigned to
each star.
\item The fluxes of all stars are summed up, with weights proportional
to the stellar initial mass function (IMF).
\end{enumerate}

SSPs are the building blocks of any arbitrarily complicated population since
the latter can be computed as a sum of SSPs, once the star formation rate
is provided.  In other words, the luminosity of a stellar population
of age $t_0$ (since the beginning of star formation) can be written as:
\begin{equation}
L_{\lambda}(t_0)=\int_{0}^{t_0} \int_{Z_i}^{Z_f} L_{SSP,\lambda}(Z,t_0-t)\,
dZ\, dt
\end{equation}
where the luminosity of the SSP is:
\begin{equation}
L_{SSP,\lambda}(Z,t_0-t)= \int_{M_{min}}^{M_{max}}SFR(Z,M,t)\,
l_{\lambda}(Z,M,t_0-t)\, dM
\end{equation}
and $l_{\lambda}(Z,M,t_0-t)$ is the luminosity of a star of mass $M$,
metallicity $Z$ and age $t_0-t$, $Z_i$ and $Z_f$ are the initial and final
metallicities, $M_{min}$ and $M_{max}$ are the smallest and largest stellar
mass in the population and $SFR(Z,M,t)$ is the star formation rate at the time
$t$ when the SSP is formed.

Our synthetic stellar population models have been extensively used in previous
works (e.g. \pcite{D+96,Spinrad+97,JPMH98,J+99}), where a detailed comparison
with other models in the literature can be found. In this work, we model
starbursts assuming that the star formation follows a Schmidt law. In
particular, we choose a star-formation law that is proportional to a power of
the amount of gas available, $SF(t)=\nu M^{n}$, where we have chosen $n=1.4$
\cite{K98} and $\nu=10$ Gyr$^{-1}$, and $M$ is the mass of gas available.
Once the star-formation law has been specified, we compute the time evolution
of the stellar spectral energy distribution (SED) using the method and the set
of SSPs outlined above. In order to compute the chemical evolution of the
starburst we have used detailed nucleosynthesis prescriptions and the
equations of chemical evolution as described in \scite{MPG98}.

\section{Dust Emission and Extinction}

In the present starburst model the SED is self--consistent from the
ultra--violet (UV) to the far--infrared (FIR), because the dust emission is
computed on the basis of the UV stellar flux, obtained from the synthetic
stellar population calculations. However, the dust extinction of the UV
stellar continuum cannot be taken into account in a rigorously
self--consistent fashion, and must rely on a specific model, that is not
constrained for example by the SED in the FIR region.  The ratio of the
``observed'' FIR and UV luminosities, $L_{FIR}/L_{UV}$, can be
$L_{FIR}/L_{UV}\ll 1$, if very little dust is present, or $L_{FIR}/L_{UV}\gg
1$, if a large quantity of dust is present and absorbs a significant fraction
of the UV continuum.  In the case of local starbursts, \scite{MHC99} find that
generally $L_{FIR}/L_{UV}>1$, with an average value of about 10, which means
that at least 50\%, and typically more than 90\%, of the UV continuum is
absorbed by dust and re-emitted in the FIR. Therefore, if it is assumed that
all UV photons are absorbed by dust and re-emitted in the FIR, the FIR
luminosity (and SED) can be computed within an uncertainty of less than a
factor of two, at least for galaxies that resemble nearby starbursts. On the
other hand, such assumption corresponds to $L_{FIR}/L_{UV}=\infty$
($L_{UV}=0$), and therefore does not allow a realistic computation of
$L_{UV}$, or alternatively, a realistic computation of the UV extinction due
to dust.  In other words, while the FIR SED (or total luminosity) can be
computed within an uncertainty of less than a factor of two, by assuming that
no UV photons escape out of the starburst, the UV extinction, or the value of
$L_{FIR}/L_{UV}$, must be estimated with some more specific model. Depending
on the model, the value of the FIR to UV luminosity ratio can be
$L_{FIR}/L_{UV}\approx 1$ or $L_{FIR}/L_{UV}\gg 1$. In the sample of local
starbursts of \scite{MHC99}, more than 1/3 of the galaxies have
$10<L_{FIR}/L_{UV}<100$, and, for any given spectral slope, the values of
$L_{FIR}/L_{UV}$ span a range of up to two orders of magnitude.

In the following we compute the FIR SED of the dust (\S 3.1), assuming that
all UV photons are absorbed by the dust (for a sufficient total amount of
dust), and then develop a novel method to estimate the dust extinction of the
UV stellar continuum (\S 3.2).

\subsection{The SED of the ``Cold'' Dust}

Our dust emission model is based on a simplified version of the Draine \& Lee
dust model \cite{DL84}, and is constructed in a manner similar to that
described by \scite{XZ89}. Xu \& De Zotti use three dust components (cold,
warm and hot dust), while we use only one component. The warm and hot
dust/molecular components are certainly the most difficult part of the
infrared spectrum to model, as they are are dominated by contributions from
PAH molecules, optically-thick dust emission from dense gas in clouds, and
warm dust emission from HII regions. However, our main interest here is to
develop a theoretical tool for the interpretation of future millimeter
observations of high-redshift objects, and therefore we avoid the
complications of computing the detailed mid-infrared SED and we are concerned
only with the spectrum at wavelengths larger than about 100 $\mu$m. In this
spectral region the emission is completely dominated by what is commonly
referred to as the `cold' dust component.

In our model, as in \scite{XZ89}, the same absorption efficiency is used for
both graphite and silicate grains.  The absorption efficiency for UV photons
is taken to be unity, and for optical and far-infrared photons is given by
equations (1) and (2) in Xu \& De Zotti (1989), which are approximations of
Draine \& Lee's results:

\begin{equation}
Q=\frac{X_{\lambda}}{1+X_{\lambda}}
\label{}
\end{equation}
where
\begin{equation}
X_{\lambda}=\frac{2\pi a}{\lambda}
\label{}
\end{equation}
for optical photons;
\begin{equation}
Q_{FIR}=0.065a\lambda^{-1.7}
\label{}
\end{equation}
for FIR photons below 130.9 $\mu$m; 
\begin{equation}
Q_{FIR}=0.065a(\lambda/0.013)^{-0.3}\lambda^{-1.7}
\label{}
\end{equation}
for $\lambda>130.9$~$\mu$m. The grain size, $a$, and the wavelength, 
$\lambda$, are measured in cm. The grain size distribution is a power 
law, 
\begin{equation}
p(a)\propto a^{-3.5}
\label{}
\end{equation}
as in \scite{MRN77}, and the number of grains per H--atom is
proportional to the metallicity of the gas, $Z$ (unlike in Xu and De Zotti's
model)
\begin{equation}
\frac{n_{gr}}{n_H}=3.5\times 10^{-10}(Z/Z_{\odot})({\rm H-atom})^{-1}
\label{}
\end{equation}
where $Z$ is the metallicity of the gas, and the factor $3.5\times10^{-10}$ is
obtained from \scite{DL84}.  In our model, the amount of dust is proportional
to the metallicity of the gas, and in fact much of the fast initial growth of
the far-infrared luminosity of the starburst is due to the build-up of
metals. We assume there is no delay between the production of metals and dust,
because the dust formation mechanism is not specified in the model.
Another major difference between the present model and Xu \& De Zotti's model
is that the intensity of the radiation field is not arbitrary, but instead is
the result of the computation of the synthetic spectrum of the galaxy.  

In order to compute the far-infrared SED of the dust, we calculate the
equilibrium temperature, $T_{eq}$, by solving the radiative equilibrium
equation for each grain of size $a$:
\begin{equation}
I Q(a) 4 \pi a^{2} \propto \int{}
B_{\lambda}[T_{eq}] Q_{FIR}(a,\lambda) 4 \pi a^{2}d \lambda
\label{dust}
\end{equation}
where $B$ is the emissivity of the black body. We then compute the 
corresponding FIR SED as a superposition of black bodies with the 
corresponding equilibrium temperatures for each grain.

We do not assume any specific dust or stellar distribution, and therefore we
cannot compute the radiation field by calculating the radiation transfer of
the UV photons through the dust. We instead assume that both the intrinsic (no
dust extinction) stellar radiation field, $I_0$, and the actual (extinguished)
radiation field to which the grains are exposed, $I$, are uniform:
\begin{equation}
I=gI_0=\frac{g L_{gal}}{4 \pi R_{gal}^2}
\label{rad}
\end{equation}
where $L_{gal}$ is the galaxy stellar luminosity, from the synthetic stellar
population calculations, and $R_{gal}$ is an effective radius, that is of the
order of the size of the galaxy, and depends on the stellar distribution.  The
factor $g$ ($g\le1$) is computed by imposing the energy balance between the
total UV stellar luminosity and the total FIR dust luminosity. This global
energy balance is necessary, because eqn. (\ref{dust}) is the energy balance
per grain, and the total FIR dust luminosity is only limited by the total
number of grains. We use $g=1$ if the dust luminosity is smaller than the
total UV stellar luminosity (the global energy balance is not imposed in this
case); we instead impose the global energy balance and determine recursively
the equilibrium temperatures and the value of $g<1$, using eqn.s (\ref{dust})
and (\ref{rad}), if the FIR luminosity for $g=1$ would be in excess of the UV
luminosity.  We do not assume any particular geometrical distribution of the
emitting dust, but simply assume that every dust grain experiences the same
uniform intensity of the radiation field, $I$, and that the dust emission is
optically thin. A better model would require a specific dust distribution, and
a radiation field computed as the result of the radiation transfer through
that dust distribution.  However, the dust distribution is highly uncertain,
and therefore we prefer to use simple assumptions, and treat the effective
radius, $R_{gal}$, as a parameter of the model. If the FIR dust luminosity is
not in excess of the UV luminosity ($g=1$), when $R_{gal}$ is decreased, the
intensity of the radiation field and the grain temperature increase. However,
$g=1$ only in the very beginning of the star formation process, when the
metallicity and the total amount of dust are extremely low.  After this
transient phase, the radiation field to which the grains are exposed is always
extinguished by the dust itself ($g<1$), and practically all UV photons are
absorbed by the dust.  As a result, the dust temperature is insensitive to
$R_{gal}$, because the energy balance yields a value of $g<1$ that keeps the
dust temperature constant (in order for the FIR luminosity to not exceed the
UV stellar luminosity). Since the parameter $R_{gal}$ has no effect in the
model, apart from an extremely short transient phase, we do not discuss it any
further. The dust temperature depends only on the total UV luminosity (the
star formation history) and on the total number of dust grains. If the stellar
luminosity increases, or the total number of grains decreases, the dust
temperature increases.  Temperature variations occur during the evolution of
the starburst (see \S 6), but they are not very large, because the dust
luminosity is proportional to the fourth power of its temperature, according
to the Stefan--Boltzmann law.

\subsection{Dust Extinction}

The interstellar reddening can be computed using reddening laws empirically
derived from observational data on nearby starbursts (e.g. \scite{C97} and
references therein). In this work we use an alternative and novel method, that
we believe is more appropriate in the interval of wavelengths typically probed
by Keck spectra of galaxies at $z\approx 3$, that is between about 1100 and
1700 \AA\, (cf Lowenthal et al. 1997).  We are able to predict the
interstellar reddening theoretically, and our main assumption is that star
formation at high redshift is a physical process similar to present-day
star-formation in our Galaxy, in the sense that most massive stars are formed
inside molecular clouds, able to extinguish significantly their far
ultra--violet (FUV) flux, until the clouds are eventually dispersed. It might
be argued that star formation in protogalactic starbursts must be more
efficient than in nearby molecular clouds, where the star formation efficiency
is very small. However, even if large molecular clouds only convert 2\% of
their mass into stars in about $10^7$ years, their gas consumption time is
$0.5\times 10^9$ years, assuming that the gas dispersed after one major star
formation episode cools in a time-scale also of the order of $10^7$
years. This gas consumption time-scale corresponds to a powerful starburst, if
applied to an entire galaxy with a large reservoir of gas. For example, a
predominantly gaseous proto-galaxy, with about $5\times 10^{10}$ M$_{\odot}$
of gas, and a gas consumption time of $0.5\times 10^9$ years, has a
star-formation rate of 100 M$_{\odot}$/yr.

Infrared surveys in molecular clouds have shown that all massive stars are
formed as members of stellar clusters, inside dense molecular cloud cores
({\it e.g.} \scite{LEDG91,LLM93}).  Although isolated star formation does
occur in some molecular clouds, such as in the Taurus--Auriga complex, this
mode of star formation does not produce massive stars. Massive stars are
formed inside dense molecular clouds, and are able to disperse their parent
clouds in about $1.5 \times 10^7$ years, thanks to stellar winds, HII regions
and supernova explosions ({\it e.g.}  \scite{H62,EL77,W79,BS80,LG82}). Because
the life-time of stars of about 15--20 M$_{\odot}$ ($Z/Z_{\odot}$=2.5--0.2
respectively) is also about $1.5 \times 10^7$ years, the FUV radiation from
stars bigger than 15--20 M$_{\odot}$ is heavily obscured during most of their
life-time by their own parent molecular clouds.  Examples of young stellar
clusters, embedded in molecular gas and dust, are listed in Table~1, together
with the fraction of their stars with infrared excess, their estimated age,
and their average visual extinction.  Stars with infrared excess are believed
to still be surrounded by their accretion discs, and so the fraction of
infrared excess sources is related to the age of the stellar cluster. Table~1
shows that it is not unreasonable to assume that molecular clouds are the main
sources of extinction for the massive stars that they form, for about $1.5
\times 10^7$ years. There is also observational evidence from external
galaxies, that massive stars in starbursts spend a large fraction of their
lives buried inside their original molecular clouds \cite{MA98}.

It is also well known that the distribution of dust in our galaxy is very
clumpy, and closely related to the distribution of dense molecular gas ({\it
e.g.} \scite{L+94}), so it is likely that dust extinction of radiation of
massive stars, due to dust distributed outside their parent molecular clouds,
is a secondary effect. Note that our assumption is in contradiction with the
assumption in \scite{C97}. There it is assumed that the effect of star
formation is to destroy most of the dust in the starburst region, and that the
main source of opacity is the dust surrounding the starburst region. According
to the assumption of our dust reddening method, stars are being formed inside
molecular clouds, that is in dense cold gas that must be associated with dust,
and such clumpy dust accounts for most of the dust in the proto-galaxy.

The global UV flux from a galaxy is dominated by the massive stars, especially
if the stellar population is young, or in the case of ongoing star
formation. Therefore, the global UV flux of a galaxy is strongly extinguished
due to the obscuration of massive stars by their parent molecular clouds,
while it is affected much less by the possibility that stars smaller than
15--20 M$_{\odot}$ are obscured by randomly intervening clouds, after the
disruption of their parent cloud.  Since the dust distribution is very clumpy,
it is to be expected that the dust surface filling fraction of the whole
galaxy is significantly less than unity, at significant extinction levels of
about $A_V\ge 0.5 $~mag.

Given the above arguments, we compute the dust extinction in high-redshift
proto-galactic starbursts, by allowing all new-born stars to remain dark for
$1.5 \times 10^7$ years. After this period of time the dark cloud is destroyed
and stars become visible. The present method ignores the possible effect of
dust as a screen acting on all the stellar population, because this is not the
dominant effect in our model. The most important property of the present
method of computing dust extinction is that it is independent of the total
amount of dust in the galaxy. The empirical method proposed by \scite{C97} is
based on the assumption that starbursts have the same stellar populations
(age, metallicity, etc.), and that their colors are only due to the amount of
dust. A larger amount of dust produces more extinction and redder colors.  The
same is true for any screen model of dust extinction. In our method, the
extinction is computed independently of the amount of dust, and no assumption
has to be made about the intrinsic stellar population, so that the properties
of the intrinsic stellar population can be constrained by comparing the model
with the observations. Results of our reddening method are shown below, where
the models are compared with the observed properties of Lyman-break galaxies.

Our way to estimate the extinction is an approximation that is valid mainly in
the FUV region of the SED, that is the region probed by Keck spectra of
galaxies at high redshift ($z>2.5$).  We do not claim that O stars are never
observed because totally obscured by their parent molecular clouds. We only
note that they are sufficiently extinguished by their parent clouds, so that
they are not the main source of FUV photons in a stellar population with a
Salpeter IMF. This can be verified empirically in the following way.  OB
stellar associations are observed in external galaxies (see for example
\scite{MP96}) and in the Magellanic Clouds
(e.g. \scite{DC91,W94,CYDTK96,TN98}). The extinction of HII regions, or of
their ionizing stars, varies typically in the range $0.2<A_V<2.0$~mag. As an
example, the extinction of HII regions in the Small Magellanic Cloud (SMC),
estimated by \scite{CYDTK96} is in the range $0.28<A_V<1.66$~mag, with an
average value $<A_V>=0.66$~mag. The same authors estimate also the extinction
of the ionizing stars, for a sample of HII regions in the SMC that partially
overlaps with the previous sample, and obtain values in the range
$0.23<A_V<0.79$, with an average value $<A_V>=0.51$~mag. As a typical
extinction we can take for example the global average using both HII regions
and their ionizing stars, and we obtain $<A_V>=0.65$~mag. In this work we will
compare the SED of our model with the SED of Lyman break galaxies (see below),
in the FUV region, between 1100 and 1600 \AA. Using a standard extinction
curve for the SMC (see \scite{CKS94} and references therein), $A_V=0.65$~mag
corresponds to $A_{1300\AA}\approx 2.5$~mag.  Therefore, HII regions in the
SMC are extinguished by dust so that typically less than 10\% of the FUV
photons emitted by their ionizing stars can escape. Since about 20\% of the
total FUV flux of a stellar population with a Salpeter IMF is emitted by stars
with mass smaller than 15--20~M$_{\odot}$ (and lifetime longer than $1.5
\times 10^7$ years), stars more massive than 15--20~M$_{\odot}$ can contribute
only for less than half of the FUV photons that escape dust absorption.  This
argument assumes that after about $1.5 \times 10^7$ years most stars will be
outside their parent molecular clouds, and not necessarily associated with HII
regions, or, in other words, that after $1.5 \times 10^7$ years $A_V\ll
0.65$~mag for the remaining B stars. The extinction measurements towards HII
regions in nearby external galaxies confirm therefore the assumption of our
extinction method.

\section{Molecular Clouds}

Molecular lines are computed using models of molecular clouds developed by
\scite{PJBN98}. The molecular emission from the whole galaxy is calculated as
the sum of the emission from single model clouds. The relative velocity
dispersion of molecular clouds in the gravitational potential of a protogalaxy
is likely to be of the order of the galaxy virial velocity, and thus much
larger than the internal velocity dispersion of typical molecular
clouds. Therefore, it is unlikely that different clouds are seen along the
same line of sight with sufficiently similar radial velocities for molecular
emission from one cloud to be significantly absorbed by the others. Here we
give only a brief description of the calculation of the molecular cloud
models, since details can be found elsewhere \cite{J97,PJBN98}.

Molecular emission lines are calculated using a non-LTE Monte Carlo radiative
transfer code (see below), which solves the radiative transfer problem for any
given density distribution, velocity field, radiation field, kinetic
temperature distribution, and chemical abundances. The density and velocity
fields are calculated by solving the three-dimensional and compressible
magneto-hydrodynamic (MHD) equations, in a regime of highly super-sonic
isothermal turbulent flows, typical of motions in Galactic molecular clouds
\cite{PJBN98}. The MHD turbulence simulations generate a large density
contrast, with the density field spanning 4-5 orders of magnitude. The density
field is very intermittent (most mass concentrates in a small fraction of the
total volume) and has a filamentary and clumpy morphology (see Fig.~1),
reminiscent of real molecular clouds.  It has been shown that the density,
velocity and magnetic fields generated by these numerical simulations are
consistent with several observational constraints
\cite{PNJ97,PJBN98,PN98,PBBJN99}, and that super-sonic and super-alfv\'{e}nic
isothermal MHD turbulence is therefore a good model for the dynamics and
structure of molecular clouds \cite{PN98}.  In this work, a run in a 128$^3$
grid of points is used, with rms Mach number ${\cal M}=15$ and Alfv\'{e}nic
Mach number ${\cal M}_A=4$, where ${\cal M}$ is the ratio of the rms flow
velocity to the sound velocity, and ${\cal M}_A$ is the ratio of the rms flow
velocity to the Alfv\'{e}n velocity. This run is the same as model Ad2 in
\scite{PN98}. The isothermal equation of state is used in the MHD equations,
and a uniform temperature is also assumed in the radiative transfer
calculations, although the Monte Carlo radiative transfer code can deal with
any temperature distribution. A range of gas kinetic temperature values
between 10 and 60 K has been used, and the cosmic background temperature has
been varied between 2.7 and 20 K, according to the assumed redshift for the
starburst.

In order to compute the luminosity of molecular lines, the abundance of the
molecules must be specified. Molecular chemistry in the context of
high-redshift proto-galaxies is discussed by \scite{FB97}, and we refer the
reader to that paper for plausible models of the evolution of important
molecules, mainly O$_2$ and CO. The abundance of O$_2$ is very uncertain, and
for simplicity we assume it is proportional to the gas O metallicity, and
equal to $1.0\times 10^{-5}$ relative to H$_2$, at solar metallicity. The CO
abundance is approximately equal to the abundance of C if O is more abundant
than C, O/C$>$1 ({\it e.g.} low metallicity or high star-formation rate),
while it is approximately equal to the O abundance if O/C$<$1, since most of
the available O is in CO \cite{FB97}. For the Sun, O/C$\approx 2$, and for the
local interstellar medium, O/C=1.2--2.5 \cite{MJHC94}. Since the chemical
evolution is computed self-consistently in our model, C and O abundances are
known at any time (see Figure~2), and we can therefore choose to set the CO
abundance equal to the C abundance if O/C$>$1, or equal to the O abundance if
O/C$<$1 (see \scite{FB97}). As can be seen in Fig.~2, with the star-formation
law assumed here it transpires that O/C$>$1 during the whole star-formation
episode, and so the CO abundance is simply set equal to the C
abundance. Finally, we have assumed that half of the total amount of gas is in
molecules, which is reasonable in the light of observational results
(e.g. \scite{YS91}). Given these assumptions, the computed luminosity of the
$^{12}$CO lines should be in error by probably a factor of approximately two
at most. However, the uncertainty in the luminosity of the O$_2$ lines is
certainly much larger, and there is the possibility that such lines could be
10 times more luminous than the value predicted by the present model.

\section{Radiative Transfer Calculations}

The non-LTE level populations were solved with a Monte-Carlo method (method B
in \scite{J97}).  The radiation field is simulated with a large number of
photon packages each representing the distributions of real photons in the
different transitions and at different Doppler shifts from the line center. In
our method the packages are always started at the edge of the cloud and
contain initially only photons from the background radiation field. During the
simulation, packages are followed through the cloud along randomly-selected
lines and the interactions between photons and gas are calculated. In each
cell, the photons that are emitted by the gas are added to the passing package
and the number of photons absorbed in the cell is counted. These counters are
later used to solve new level populations before the simulation continues with
a new set of photon packages. As a photon package exits the cloud the number
of photons is recorded. The recorded values represent the luminosity of the
cloud in the different transitions integrated over the whole cloud surface.

For the calculations the cubic model cloud was divided into 64$^3$ cells where
the density and velocity values were interpolated from the original 128$^3$
data-cube that is the result of the MHD calculations. For each cell the
intrinsic non-thermal velocity dispersion was estimated from dispersion of the
velocity values in the original data-cube. Together with the thermal line
broadening this defines the local absorption and emission profiles for the
radiative-transfer calculations. The reduction of the linear resolution by a
factor of two (from 128 to 64 grid points) is not expected to affect the
results of the radiative-transfer calculations significantly. The density
variations are still resolved and most cells remain optically thin. The
limited resolution might slightly reduce the emission from transitions between
high-energy levels ($J\sim$10) that are populated only in the rarest, densest
cells. No such effects are expected for the lower-level transitions which are
emitted by much larger regions. This has also been confirmed in previous
calculations with similar models.

The model cloud has a linear size of 20\,pc and a mean density of
200\,cm$^{-3}$. The total range of densities extends from just under
1\,cm$^{-3}$ to 1.25$\times$10$^4$cm$^{-3}$. The critical density of the
CO(1--0) transition, $n_{\rm cr}$=1.8$\times$10$^3$cm$^{-3}$, is exceeded in
only some 2\% of the cloud volume and most of the CO is clearly sub-thermally
excited.

The background radiation field consists of CMB at the corresponding redshift
({\it e.g.} models discussed below have $T$=15\,K corresponding to $z=$4.5)
and an additional stellar and dust radiation-field component, calculated from
the starburst model for the selected epoch. The background is generally
dominated by the CMB in the frequency range relevant for the excitation of CO.
At $4.8\times 10^7$ years the background intensity due the starburst exceeds
temporarily that of the CMB at high frequencies and the CO transitions between
levels above $J\approx8$ are subjected to a much hotter background. The number
of molecules in these high-excitation levels is, however, very small and no
effect on the intensities of the lower transitions is seen.

The kinetic temperature of the cloud was assumed to be uniform. For example,
assuming $T_{\rm kin}=$60\,K, the population of CO levels $J\ga$10 remains low
even when molecules are thermalized. The number of excitation levels included
in the radiative transfer calculations was 13 and the population of the
highest levels was found to be negligible even in the densest regions. These
high-excitation levels may still be important for the cooling of
optically-thick clouds when the lower transitions become saturated.  However,
they will have very little effect on the populations on levels $J<$10.  In our
case the mean optical depths through the cloud are, for example for the CO
transition $J=8-7$, already well below 1.0 and it is clear than no noticeable
errors are caused by the exclusion of levels $J\ge13$. The largest optical
depths were found for transitions $J=4-3$ and $J=3-2$ where the average values
were close to 20. Most of the CO is therefore shielded from the external
radiation.

The results were quite similar for all epochs simply because the external
radiation field was always dominated by the CMB (since in this work we
concentrate on protogalaxies at $z > 3$). The mean excitation temperatures
varied in all cases from some 33\,K for the transition $J=1-0$ to about 21\,K
for $J=8-7$. The excitation temperatures of the higher transitions are
somewhat uncertain since in most of the cloud these levels are simply not
excited. For the lowest transition the mean excitation temperature for cells
with $n<100$cm$^{-3}$ was in all cases close to 22\,K while the corresponding
number for cells with densities higher than 100\,cm$^{-3}$ was close to
37\,K. The excitation does not depend, of course, only on the local gas
density but on the surrounding cells and the general position inside the
cloud.

We have computed lines for other relevant molecules (O$_2$ etc) but none of
them are as luminous as the $^{12}$CO lines. In particular, the O$_2$ lines,
the next more luminous after the $^{12}$CO ones, are only 10\% as bright as
the latter. It is possible, however, that the O$_2$ lines are more luminous
than predicted here, due to the large uncertainties in the O$_2$ abundance, as
commented above.

\section{Results: Time evolution of the spectral energy distribution}

In Figure~3 we show the time evolution of the SED for a
starburst of $10^{11}$ M$_{\odot}$, where reddening has been computed with the
method described above; the $^{12}$CO lines are also displayed. The last 
panel (age=0.3 Gyr) shows the SED without dust, as it could be for example
if dust were to be removed due to the energy deposited by supernovae. We
have assumed that the starburst forms at $z=4.5$, with $T_{\rm CMB}=15$~K; 
we have also assumed that the kinetic temperature of the molecular gas
is the same as the average temperature of the dust, $T_{\rm kin,CO}=30$~K. 
$^{12}$CO lines are always in emission and are 10 times stronger
than O$_{2}$ lines. Although we have chosen a particular star formation law
that may be unrealistic for high mass objects \cite{JFDTPN99}, the important
point is that we can produce self-consistent models from the UV to the far-IR
and mm using the detailed radiative transfer calculations and the new
reddening method.

Although it is a bit difficult to appreciate from the plots in Figure~3, in
our starburst model the dust temperature decreases slightly with time, from an
average temperature of about 40~K to about 25~K. The dust temperature in our
model is only affected by the stellar UV radiation field, computed with the
synthetic stellar population calculations, and by the total amount of dust.
The most important effect is certainly the variation of the total amount of
dust, due to the increasing metallicity. When the amount of dust increases,
the UV radiation field to which grains are exposed decreases, as it is
extinguished by the dust itself (the value of the parameter $g$ in eqn.
(\ref{rad}) decreases), and therefore the grain temperature decreases.  As an
example, we consider two high--redshift radio--galaxies: 4C41.17 and 8C1435.
Figure~4 displays the measurements obtain for these two objects at FIR
wavelengths by Archibald (1999) (private comm. PhD Thesis, University of
Edinburgh) as diamonds. We have also plotted our starburst model (continuous
line), appropriately redshifted. We find that 4C41.17 can be well fitted by
our starburst model with an age of about $10^8$~yr, which corresponds to a
grain temperature distribution between 20 and 28 K, while 8C1435 is better
fitted with an age of about $10^7$~yr, which corresponds to grain temperature
between 45 and 32 K. A maximum likelihood method for a grey body, where
$\beta$ and the dust temperature are the free parameters, yields (for
$\beta=2$) to temperatures of 25 K for 4C41.17 and 40 K for 8C1435, in
agreement with our model.

In Figure~5 we show the time evolution of the $^{12}$CO luminosity for several
transitions. The luminosity grows for about 0.1--0.2 Gyr, because of the
increase in the metallicity and therefore in the abundance of CO molecules,
while after 0.2 Gyr the luminosity decreases due to the gradual consumption of
gas into stars. The most luminous transitions at 0.1 Gyr are the $J=5-4$ and
the $J=6-5$. If only two of the $^{12}$CO luminous transitions are detected in
high redshift starbursts, their relative frequency will give an extremely
precise measure of the redshift of the starburst.

\section{Comparison with observations}

\subsection{Lyman--break galaxies}

The most important optical window into the high-redshift universe comes from
the discovery of starbursts using the Lyman-limit imaging technique (see
\scite{SPH95} for origins of this method and \scite{S+98} and references
therein for the latest results). The method is conceptually very simple: the
flat spectral energy distribution of starbursts red wards of 912 \AA\, and the
increasing opacity of the intergalactic medium at high redshifts
(\pcite{YP94}), produce a distinctive `step' in the observed optical spectrum
of star-forming galaxies at high redshift. Therefore, using filters selected
to sample the continuum blue wards and red wards of the 912 \AA\,
discontinuity at a given redshift, one can image large fields and successfully
discover a population of high-redshift star-forming galaxies by the lack of
light blue wards of 912\AA\, coupled with a clear detection in redder
filters. Spectroscopically, the star-forming galaxies found by this technique
form a rather homogeneous sample, and so rather general conclusions regarding
dust extinction and age of such objects can be drawn from the detailed
analysis of the spectra of a few such galaxies.

Figure~6 shows the predicted time evolution of the spectral energy
distribution of simple stellar populations (see \S 2), that are the basic
ingredients of our model. A simple stellar population (SSP) is a population of
stars formed all at the same time and with the same metallicity.  The four top
panels of Figure~6 show the case of SSPs with four different values of
metallicity, at three different ages, 10, 20 and 30 Myr.  We have only plotted
the synthesized spectra over the wavelength range $1200 < \lambda < 1800$
\AA\, since, by virtue of their selection, this is the spectral region of
interest for comparison with the observed optical spectra of Lyman-limit
galaxies. In the two lower panels of Figure~6, the SSPs of with $Z=0.2
Z_{\odot}$ and $Z=Z_{\odot}$ are shown again (only for the two ages of 10 and
20 Myr), but the SED are normalized at 1800 {\AA}. The four upper panels are
therefore useful to appreciate the dimming of SSPs with time, while the two
bottom panels show better the variation of the spectral slope.  The continuum
in this wavelength interval can be well-approximated by a power law ($F_{\nu}
\propto {\lambda}^{\alpha}$); ${\alpha}$ changes with time and metallicity
from -1.5 to +1.6. It can also be seen that the rate of dimming of SSP due to
aging is essentially independent of metallicity (four upper panels of
Figure~6), which is the reason why our extinction method, applied to the self
consistent starburst model, yields a roughly constant value of the
$L_{FIR}/L_{UV}$ ratio.

We now compare the predictions of our starburst model, with continuous star
formation and self--consistent chemical evolution, with the Lyman--break
galaxy 1512--cB58 (\pcite{PSADG99}), which is one of the reddest among the
Lyman-break galaxies ($\alpha=+1.3$). Most of the absorption features seen in
1512--cB58, plotted in Figure~7, are interstellar features; since we are not
aiming at modeling those, we judge the success of our model only by the extent
to which it can accurately reproduce the continuum FUV SED of this galaxy. Any
successful fit between the stellar features of our model and the observed
spectrum of 1512--cB58 must be regarded as coincidental (but see below). To
stress this difference, we choose to plot the SED of 1512--cB58 at a higher
resolution than the SED of our model. In Figure~7 we plot the SED of our model
at 0.4 Gyr and solar metallicity (lower panel), and at 0.8 Gyr and three times
solar metallicity (upper panel) (metallicity grows with time in our model).
We also show, as dashed lines, the SED of the same model, but without any
reddening correction, that is no 15 Myr delay for the stars to appear out of
their parent giant molecular clouds. In the lower panel we also show as dotted
line the case with a 30 Myr delay.  It can be seen in Figure~7 that the model
without reddening is always too blue, even with a very high metallicity, to
reproduce the observed spectrum.  The model with the reddening computed as
explained in \S 3.2, with a 15 Myr delay, reproduces the shape of the UV
spectrum very well, with a more reasonable metallicity ($Z_{\odot}$). Although
the model with a delay of 30 Myr could fit the observed spectrum if a
metallicity lower than solar were adopted, there is no observational support
in favor of such a large delay.  Finally, a model with a metallicity as low as
0.1 $Z_{\odot}$ (not shown in Figure~7) does not provide a good fit to the
observed spectrum, unless we assume a level of dust extinction corresponding
to molecular clouds obscuring their stars for about 0.2 Gyr, which cannot be
justified with observational data.

As stated above, 1512-cB58 has a UV spectral index of $\alpha=+1.3$. The
Lyman-limit population as a whole displays a range of UV spectral indices
$\alpha$ between 0 and 1.5. The lower limit of this range can be understood
with our model, because $\alpha\approx 0$ for an age of about 15~Myr, and
$\alpha< 0$ for lower ages. Since our extinction method is based on the
argument that molecular clouds extinguish very heavily (at least 90\%) of the
FUV stellar continuum for about 15~Myr, galaxies with $\alpha< 0$ are expected
to be rather dim in the FUV and therefore hard to detect.  The upper limit for
$\alpha$ is also explained in our model, because we find that our reddening
method predicts that spectral slopes larger than 1.5 are extremely unlikely,
since they require a metallicity of 10 $Z_{\odot}$. Our model thus naturally
explains why Lyman-break galaxies should display a limited range of values for
$\alpha$. In contrast, we emphasize that in the dust screen model a larger
range of values of $\alpha$ is in principle allowed, since the value of
$\alpha$ depends mainly on the assumed total amount of dust.

One more interesting conclusion can be drawn from Figure~7. Since all the
absorption features in the observed spectrum are known to be due mainly to the
interstellar medium, with a small stellar contribution, the model with high
metallicity can in fact be ruled out simply because it already provides
stellar features that are stronger than the features in the data.  A much
better fit is obtained by the $Z_{\odot}$ model both to the continuum shape,
and to the lines ({\it i.e.} weaker absorption lines than observed).

\subsubsection{Evidence for Very Massive Stars in the FUV Spectrum of Lyman-Limit Galaxies}

In the method to compute FUV dust extinction presented in this work, it is
assumed that stars are invisible during the first 15 Myr of their life. This
is just an approximation. As explained in \S 3.2, stars are expected to be
extinguished at FUV wavelengths by at least 2-3 magnitudes during the first 15
Myr of their life, rather than to be totally invisible. However, a prediction
of our extinction method is that FUV spectral features of stars more massive
than 15-20 M$_{\odot}$, should therefore be heavily extinguished, and hardly
visible in the starburst spectrum, together with their FUV continuum, because
the lifetime of such stars is about 15 Myr.

Recently, unprecedently high quality data of one Lyman-break galaxy,
1512--cB58, has been obtained by \scite{PSADG99}, the spectrum of which shows
classic P--Cygni profiles arising from several high-ionization species such as
C~IV and N~V. \scite{PSADG99} have concluded that the C~IV P--Cygni profile
provides evidence for the presence of very massive stars (M $>
30$M$_{\odot}$).  However, by comparing spectral synthesis models to the
observed C~IV P--Cygni profiles, \scite{PSADG99} actually show that a standard
Salpeter IMF fits well the emission component of the P--Cygni profile, but
very poorly the absorption component, while an IMF with only stars less
massive than 30 M$_{\odot}$ does not provide a good fit to the emission
component, but a much better fit to the absorption. In other words, what they
really prove is that their spectral synthesis models are inadequate to
describe fine details of the SED of the galaxy 1512--cB58, such as P--Cygni
features.  \scite{PSADG99} also find that most absorption lines in the
spectrum of 1512--cB58 originate mainly in the interstellar medium, that is
clearly expanding at a velocity of a few hundreds km/s. The same C~IV
absorption is dominated by interstellar medium absorption. It is therefore
hard to believe, especially in the presence of large interstellar medium
outflows, that the absorption components of P--Cygni profiles are so strongly
affected by the interstellar medium, while the emission components have purely
stellar (or stellar wind) origins.  It is also not clear whether 1512--cB58 is
at all representative of the rest of the Lyman-break galaxy population, since
many Lyman--limit galaxies (e.g. \scite{Lo+97} and Steidel et al. sample) do
not show any P-Cygni profiles (although that might be in part a consequence of
the inferior spectral resolution and signal--to--noise). Models of C~IV
P--Cygni profiles of large interstellar medium outflows in starbursts are
needed in order to quantify the possible contribution of the interstellar
medium not only in absorption, but also in emission.

\subsection{Dust emission from high-redshift galaxies}

Using the extinction method presented in \S 3.2, we can compute the ratio
$L_{FIR}/L_{UV}$, where $L_{FIR}$ and $L_{UV}$ are the ``observed'' FIR and UV
luminosity of the starburst (see beginning of \S 3). However, this ratio is
highly dependent on the wavelength interval over which the flux is integrated,
or, in the case of generalized fluxes of the type $\lambda f(\lambda)$, on the
wavelength $\lambda$ at which the flux density $f(\lambda)$ is measured,
because the FIR emission is strongly peaked around 100~$\mu$m.  Since the FIR
luminosity is expected to be approximately equal to the UV luminosity (see
beginning of \S 3), the ``observed'' $L_{FIR}/L_{UV}$ ratio is mainly a
function of the ratio between the intrinsic (no extinction) and the
``observed'' (extinguished) UV luminosities. We therefore define the
correction factor as the ratio between the intrinsic and the reddened fluxes
at 1600~{\AA}, $(f_0/f)_{1600}$, and use it in alternative to the
$L_{FIR}/L_{UV}$ ratio.

The FUV luminosity absorbed by dust corresponds roughly to the FUV luminosity
of all stars in the first 15 Myr of their life, during which at least 90\% of
their FUV photons are absorbed by grains.  For a steady starburst lasting
substantially longer than 15 Myr, this leads to a robust prediction that,
$(f_0/f)_{1600}\approx 6$, with no strong time variations (as explained in the
discussion of Figure~6, in \S 7.1), which is also the correction factor for
star formation rate estimates based on the observed FUV flux. This result
depends little on metallicity, provided that the metallicity is sufficiently
high for significant quantities of dust to have formed ($Z\ga 0.01
Z_{\odot}$). If the starburst is younger than 15 Myr, the correction factor is
$(f_0/f)_{1600}> 6$, and can be as large as $(f_0/f)_{1600}\sim 20$, since
very young OB stars (few Myr) can easily be extinguished by their parent
clouds up to a few magnitudes in the FUV.  Nearby starbursts, or starbursts at
intermediate redshift, are likely to be forming stars discontinuously, mainly
during periods of interactions with other galaxies, or merging episodes. In
that case, it is possible that some nearby starbursts are observed during a
major episode of star formation that has not started longer than 15 Myr ago,
and the correction factor could be larger than five. However, it is likely
that most galaxies at high redshift are forming stars with more continuity
than nearby starbursts, because the rate of merging and interactions at high
redshift is much higher than at low redshift, and because of the larger amount
of gas available to form stars at high redshift. Our model predicts therefore
that the correction factor $(f_0/f)_{1600}\approx 6$ for high redshift
starbursts, such as Lyman-limit galaxies. A factor much larger than 6 for a
significant fraction of the population of Lyman-limit galaxies would be
possible only if their starbursts were very intermittent episodes, never
lasting much longer than 15 Myr.

If we define the ``observed'' $L_{FIR}/L_{UV}$ ratio as $L_{FIR}/L_{UV}\equiv
(f_0/f)_{1600}-1$, then our model predicts that the FIR (rest--frame) flux of
Lyman-limit galaxies is about 5 time larger than their UV flux:
$L_{FIR}/L_{UV}\approx 5$.  The $L_{FIR}/L_{UV}$ ratio can be related to the
values of the spectral slope, $\alpha$. In our model $L_{FIR}/L_{UV}$ is
independent of $\alpha$, for starbursts older than 15 Myr. For younger
starbursts $L_{FIR}/L_{UV}$ can be much larger, but since our extinction model
does not work in that case, we cannot define $\alpha$ there. This is not an
important limitations of the model, since, as discussed in \S 7.1, so young
starbursts are extinguished so much (at least 2-3 mag in the FUV), that they
are certainly very poorly represented in Lyman-limit galaxy samples.  In other
words, our theoretical relation between $L_{FIR}/L_{UV}$ and $\alpha$, is
limited to $\alpha \ge 0$, as real samples of Lyman-limit galaxies are. Our
result is in stark contrast to the prediction of screen models, for which both
$\alpha$ and $L_{FIR}/L_{UV}$ are directly related to the total amount of
dust. In Figure~8 we compare a screen model \cite{Peacock+99} with the
prediction of our extinction method.

Some evidence for $L_{FIR}/L_{UV}\approx 5$ is emerging in the comparison of
the faint (below 2 mJy) SCUBA sources, with their HDF counterparts
\cite{Peacock+99}.  Given that 8C1435 and 4C41.17 appear to be extremely
massive starbursts which have been in progress for some time (Dunlop et
al. 1994; Ivison et al. 1998) it is interesting to consider whether their
observed properties are also consistent with $L_{FIR}/L_{UV}\approx 5$.  Based
on its observed 850$\mu m$ flux density of 12 mJy (Dunlop et al. 1994) our
model predicts that 4C41.17 should have an observed I-band magnitude of I=
22.2. This is in excellent agreement with its observed I-band magnitude of
I=22 $\pm$ 0.5 (Chambers et al. 1990), which implies $L_{FIR}/L_{UV}\approx 5$
also for 4C41.17. It is not surprising therefore that Dey et al. (1999) have
recently demonstrated that the ultraviolet SED of 4C41.17 looks exactly like
that of Lyman-limit galaxies, only enhanced by a factor of $\simeq 100$, as if
4C41.17 were an extremely massive (or massive collection of) Lyman-limit
galaxy(ies) (apart from its extremely luminous radio emission).  Similar
arguments appear to apply to 8C1435, although in this case the
optical-ultraviolet SED of the radio galaxy has not been studied in as much
detail, and the match between predicted and observed $I$-band magnitude is
less optimal (I=23 is predicted whereas 8C1435 is actually 0.5 magnitudes
fainter than this --50\% fainter than the luminosity predicted by our model).

We can now apply these predictions to assess the plausibility of some of the
optical identifications which have been suggested for the brightest
sub-millimeter sources detected by Hughes {\it et al.} in their deep 850$\mu
m$ SCUBA survey of the Hubble Deep Field \scite{H+98}.  Hughes {\it et al.}
discovered at least 5 clear sub-mm sources with $S_{850} > 2$ mJy and
attempted to assess the statistically plausibility of the possible association
of these sub-millimeter sources with optically detected galaxies already known
in the HDF. In fact, due to the large angular size of the SCUBA beam, in most
cases several optical candidates are available and this process yielded 1, or
at most 2, reliable optical identifications. Despite this, the
sub-millimeter--infrared colors of the sources, combined with the lack of
possible low-redshift optical counterparts, led Hughes {\it et al.} to
conclude that most (4 out of 5) of these sources lie at $z \ge 2$. Our new
models allow us to verify whether a proposed optical identification appears
physically plausible given, for example, the inferred sub-millimeter to
ultraviolet luminosity ratio, and the colors of the optical identification.

We choose to concentrate on the brightest sub-mm source, HDF850.1, because for
this source the photometric data are of sufficient quality to indicate that it
almost certainly lies at $z > 2$ and most likely at $z > 3$ (note that both
flux ratios (450/850 and 850/1350) yield the same answer in our models
(c.f. \scite{H+98}). We can use our models to explore whether the optical
galaxy (3-577.0), tentatively suggested by Hughes {\it et al.} as the most
likely optical counterpart of HDF850.1, really does have properties consistent
with those of a starburst galaxy capable of producing 7 mJy of flux at 850$\mu
m$.  We have done this as follows. Adopting the optical galaxy 3-577.0, which
has a tentative redshift of $z = 3.36$, as the correct identification, we have
computed our starburst model at $z = 3.36$ and deduced age (and hence
formation redshift) that best fits the measured optical colors of 3-577.0
(from the optical HDF \cite{W+96}). The result is that we can obtain an
excellent fit to the $u-b$, $b-v$ and $v-i$ colors of 3-577.0, for a starburst
age of about 0.6 Gyr. Our extinction model, that requires an age of more than
15 Myr, can be applied, and predicts $L_{FIR}/L_{UV}\approx 5$, while the
optical identification 3-577.0 provides a value $L_{FIR}/L_{UV}\approx 30$. We
conclude therefore that either the optical identification is incorrect, or the
starburst is younger than 15 Myr. Another possibility is that the sub-mm
source HDF850.1 could be a collection of physically unrelated sub-mm sources,
possibly at different redshifts, due to the finite resolution of the SCUBA
map. Since the sub-mm flux from dust at redshift between 1 and 10 is hardly
sensitive to redshift, it is possible that many sub-mm sources are detected
along the line--of--sight at different redshifts, while only the closest or
brightest optical counterparts can be seen in the HDF.

\subsection{Nearby starbursts}

Local starbursts may, in principle, be very similar to protogalaxies, since
they have, like protogalaxies, large reservoirs of gas which are rapidly
turned into stars.  It is therefore worth comparing our models with the
observational properties of local starbursts.  This comparison is interesting
because local starbursts have been used as the main laboratory in studying
dust extinction of stellar light in galaxies (\pcite{CKS94,C97}), and results
of such studies are often applied to high redshift galaxies
(e.g. \scite{PKSDAG98}) to infer the properties of their intrinsic
(unreddened) stellar population.

Local starbursts have been used to infer a universal extinction curve, by
assuming that the intrinsic stellar population of the sample as a whole is the
same (\scite{C97}). Under this assumption, the different UV colors are believed
to be due to different amounts of dust extinction, and not to differences in
the ages of the stellar populations. The assumption is supported by the fact
that i) some synthetic stellar population models at {\it constant solar
metallicity} (BC96, LH95 (\pcite{L+96})) do not seem to fit the observed
colors of local starbursts if dust extinction is not accounted for, and ii)
the color of the stellar continuum correlates with the dust extinction in the
nebular lines (the redder the stellar light the more dust extinction in the
nebular line).

However, synthetic stellar population models that take into account
self--consistently the chemical evolution of the starburst, and which include
the effects of dust extinction, can indeed fit the colors of local starbursts
and explain them as a metallicity (age) effect (see below). Moreover, the
correlation between the colors of starbursts and the extinction of nebular
lines can be understood as a consequence of the fact that older starbursts can
have more dust than young starbursts, because their metallicity is larger, and
because their stars have had more time to form dust grains.

Adopting a single, intrinsic spectrum for all starbursts is in apparent
contradiction with the range of $J-H$ and $H-K$ colors of the
starbursts. These infrared colors are not significantly affected by dust
extinction, and prove that the stellar population in the starbursts have ages
that span from about $10^7$~yr for the bluest and youngest galaxies, to about
$0.6\times 10^9$~yr for the reddest and oldest ones (see below). The oxygen
abundances of the same sample also span a range from one fourth solar to twice
solar. It therefore seems very unlikely that a single composite stellar
spectrum, modified by dust, is a good assumption for the SEDs of nearby
starburst galaxies.

We have compared our starburst models with the $J-H$ and $H-K$ infrared
colors, with the 0.16 $\mu$~m to 2.18 $\mu$~m flux density ratio, and with the
estimated metallicity of local starbursts found in \scite{C97}. We find that
the range in the infrared colors and in the metallicity are reproduced by the
model, in the age interval $15 \times 10^7-0.6\times 10^9$~yr (Fig.~9), which
means that some local starbursts have extremely young stellar populations,
while some are 60 times older.  The youngest starbursts might have an older
stellar population that does not dominate the stellar continuum, but 
the stellar spectrum is still dominated by populations with
very different ages in different starbursts. We also find that our model
reproduces the observed $F(0.16)/F(2.18)$ flux density ratio, in exactly the
same age interval ---  the colors of the stellar continuum are therefore
mainly an effect of age. Notice that a starburst model that does not describe
self--consistently the chemical evolution (for example, a model with a fixed
metallicity) cannot account for the observed range of values of infrared and
UV colors, and of course for the range of metallicity. The aging of the
stellar population alone does not produce a sufficient change in colors, or
in the spectral slope. The change in colors is due to the
fact that in a self consistent model, such as the one described here, the
metallicity has to grow with age, and so the stellar continuum becomes redder.

In our model, older starbursts are redder than younger ones because they have
a higher metallicity, not because they have more dust.  Although it is
probably true that older starbursts have more dust than younger ones (and this
explains the trend of nebular line extinction with stellar colors), increasing
the amount of dust simply makes the approximation of our method even better,
because most dust is inside clumpy molecular clouds, and a larger amount of
dust means that stars embedded in their parent molecular clouds are even
better hidden and do not contribute significantly to the UV continuum. In our
model the amount of dust determines how good our approximation is, but it is
not the main reason for the colors of the starbursts.

The scenario that dust and stellar extinction concentrate in molecular clouds
does not contradict the result of \scite{CKS94}, who find that dust extinction
of nebular lines is dominated by foreground, rather than internal, dust. In
our model massive stars are embedded in their parent molecular cloud together
with the HII region around them, until the HII region can completely disperse
the cloud. Most dust inside the HII region is probably destroyed, and so the
emission lines from the ionized gas must be absorbed mainly from the dust
distributed in the remaining part of the molecular cloud that has not been
ionized yet. This dust is between the HII region and the observer; hence
foreground dust, although concentrated in the molecular cloud, causes the
extinction of the nebular lines.  This well agrees with the facts that clumpy
foreground distributions are needed to interpret the observations
(\pcite{CKS94,C97}), and that dust in molecular clouds is known to have a very
clumpy distribution (\pcite{L+94,PNJ97}).

We conclude that the adoption of a single SED representing the population of
starbursts as a whole is likely erroneous. Moreover, our new assumptions allow
a better comparison between theoretical models and observations, because the
free parameter describing the total amount of dust becomes almost irrelevant
for the colors of the starbursts.

\subsection{Elliptical galaxies at high redshift}

An interesting issue to address is how the starburst model presented in \S 2
would evolve after the burst is finished, and how it compares with the reddest
known elliptical galaxies at $z\sim 1.5$ \cite{D+96,Spinrad+97,D98}. For this
purpose, we use a starburst model of $5\times 10^{11}$ M$_{\odot}$, whose star
formation activity last for 0.7 Gyr, and compare it with the observed SED of
53w069 \cite{D98}(the reddest known galaxy at $z \sim 1.5$). As judged by
minimizing $\chi -$squared we find that the best fit is achieved at an age of
4.0 Gyr (upper panel of Figure~10) --- we remind the reader that 53w069 is at
$z=1.47$ and thus that an age of 4 Gyr implies either that an Einstein-de
Sitter Universe is incorrect, or that $H_0<50$ km s$^{-1}$ Mpc$^{-1}$. This
age is slightly older than the one found in \cite{D98} using a model where the
burst takes place in an infinitesimal time and assuming solar metallicity for
the whole population. In contrast, here we are using a model where the
metallicity is building self-consistently with the star formation rate and
chemical evolution models discussed in \S 2.

In order to investigate further the origin of this slight ``aging'' relative
to the model with a single metallicity, we have plotted in the lower panel of
Figure~10 the contribution to the theoretical SED from stars born with
$Z<Z_{\odot}$ (continuous line) and $Z>Z_{\odot}$ (dotted line). Not
surprisingly the low-metallicity stars dominate the UV part of the spectrum;
short ward of 2900\AA\ they contribute virtually all of the galaxy light,
despite the fact that they comprise only 20\% of the mass of the galaxy. On
the other hand, for the part of the spectrum red-wards of 3000 \AA \, both
populations contribute equally. The important lesson to be learned from this
modeling is that attempting to follow the build-up of metals is of crucial
importance for interpreting the spectra of galaxies. In the case of 53W069,
the adoption of a high-metallicity population, without taking account of the
previous low metallicity population, albeit small, would lead to a gross
under-estimate of its age.

We now consider the color evolution of our starburst model with redshift.  For
simplicity we will discuss only the case of an Einstein-de Sitter Universe
($H_0=50$ km s$^{-1}$ Mpc$^{-1}$) and will assume that the starburst is formed
at $z=5$. Six colors are plotted versus redshift in Figure~11. Looking at the
evolution of $R-K$ it is clear that the colors expected for the 53W galaxies
\cite{D+96,Spinrad+97,D98} are well fitted. Although dependent on the exact
duration of the burst, the reddest colors are expected at $1.2 < z < 2$, as it
is observed \cite{D98}. Note that this is completely the opposite to what is
predicted in the unrealistic model where the burst is assumed to take place in
an infinitesimal time, since then the reddest colors are expected at
$z>4$. This effect produced by early star formation is also clear in the
evolution of $V-K$ and $I-K$. In the $V-K$ versus $z$ plot, we also show the
limit found by \scite{Z97} for the lack of red faint objects in deep
surveys. The largest value of $V-K$ predicted by our starburst model is about
1 mag bluer than the value predicted by Zepf, using a model with instantaneous
star formation, and assuming solar metallicity.  In a recent paper
\cite{JFDTPN99}, we have studied with a detailed hydro-dynamical model the
evolution of single-collapse objects at high-redshift and shown that such
models are always much bluer and not as faint as predicted with starburst
models where the duration of the burst is assumed to be infinitesimal
\cite{Z97}. In \scite{JFDTPN99} a much more detailed comparison with
observations is given, along with predictions of the space-density of faint
red objects, and we refer the interested reader to that paper for further
details.

As expected, these high-redshift starbursts end up at $z=0$ with spectral
properties characteristic of E/S0 galaxies and bulges. This point is
illustrated in Figure~12, where we have compared the theoretical SED at $z=0$
with an average SED for E/S0 and Sa galaxies from the Kennicutt catalog
\cite{K92}. One would therefore be tempted to conclude, solely from their
stellar content, that high-redshift starburst activity is the ``engine''
behind the formation of bulges and elliptical galaxies.
                                                 
\section{Conclusions}

We have modeled the spectral energy distribution of proto--galactic starbursts
at high redshift, by calculating self--consistently synthetic stellar
populations, chemical evolution, dust emission and extinction, and molecular
emission. Dust extinction is calculated in a novel way, that is not based on
empirical calibrations of extinction curves, and molecular emission is
computed using three--dimensional magneto--hydrodynamic simulations of highly
super--sonic turbulence in molecular clouds, and a non--LTE radiative transfer
Monte Carlo code.

The main result of the present work is that a model of a 
proto--galactic starburst at high redshift can account for
observed properties of:

\begin{itemize}
\item SCUBA/HDF sources and their optical counterparts;
\item Lyman--limit galaxies;
\item High--redshift radio--galaxies; 
\item High redshift elliptical galaxies;
\item Galaxies detected in deep surveys (their colors);
\item Nearby elliptical galaxies.
\end{itemize}

This might be an indication that nearby elliptical galaxies, high redshift
`elliptical' galaxies, and Lyman--limit galaxies are the same objects observed
at different epochs of their evolution, and are all formed by rather short
($<$1~Gyr) starbursts, during which most of the proto--galactic baryons are
turned into stars, at redshifts $2.5<z<6$ (or larger than 6), and the star
formation process starts probably at $z\ge 5$, for most of these
galaxies. Because the star formation process is likely to be coeval with the
collapse of galactic halos, which can last about 0.5 Gyr for typical galactic
masses and formation redshifts, it is not easy to use the present models to
discriminate between a cold dark matter hierarchical (merging) galaxy
formation, and a monolithic galaxy formation scenarios. The two scenarios are
probably expressions of two different points of view of the same process and
can be reconciled with each other.

Another important result is that according to our assumptions and our method
of calculating the dust extinction, it is possible to quantify the effect of
dust on the UV flux of the proto--galaxy without any free parameters. We find
that a correction factor of 6 to the FUV flux must be applied for values of
the UV slope $\alpha$ between 0 and 1.5, while the correction factor can be
much larger than 6, for starbursts younger than 15 Myr, for which our
extinction model cannot be applied. We also predict that young starbursts with
$\alpha > 1.5$ will be extremely rare.

We thank the anonymous referee for many useful comments.  MJ acknowledges the
support of the Academy of Finland Grant no. 1011055.

\clearpage


\begin{thebibliography}{{Calzetti}, {Kinney} \& {Storchi-Bergmann}<1994>}
\bibitem[{Aguirre} \& {Haiman}<1999>]{AZ99}{Aguirre} A., {Haiman} Z.,
  1999.\newblock {\rm astro-ph}, {\rm 9907039}.
\bibitem[{Blitz} \& {Shu}<1980>]{BS80}{Blitz} L., {Shu} F.~H., 1980.\newblock
  {\rm ApJ}, {\rm 238}, 148.
\bibitem[{Calzetti}, {Kinney} \& {Storchi-Bergmann}<1994>]{CKS94}{Calzetti} D.,
  {Kinney} A.~L., {Storchi-Bergmann} T., 1994.\newblock {\rm ApJ}, {\rm 429},
  582.
\bibitem[{Calzetti}<1997>]{C97}{Calzetti} D., 1997.\newblock {\rm AJ}, {\rm
  113}, 162.
\bibitem[{Caplan} {\rm et~al.}<1996>]{CYDTK96}{Caplan} J., {Ye} T., {Deharveng}
  L., {Turtle} A., {Kennicutt} R., 1996.\newblock {\rm A\&A}, {\rm 307}, 403.
\bibitem[{Deharveng} \& {Caplan}<1991>]{DC91}{Deharveng} L., {Caplan} J.,
  1991.\newblock {\rm A\&A}, {\rm 259}, 480.
\bibitem[{Draine} \& {Lee}<1984>]{DL84}{Draine} B.~T., {Lee} H.~M.,
  1984.\newblock {\rm ApJ}, {\rm 285}, 89.
\bibitem[{Dunlop} {\rm et~al.}<1996>]{D+96}{Dunlop} J., {Peacock} J., {Spinrad}
  H., {Dey} A., {Jimenez} R., {Stern} D., {Windhorst} R., 1996.\newblock {\rm
  Nature}, {\rm 381}, 581.
\bibitem[{Dunlop}<1997>]{Dunlop_sfr98}{Dunlop} J., 1997.\newblock {\rm
  astro-ph}, {\rm 9704294}.
\bibitem[{Dunlop}<1998>]{D98}{Dunlop} J., 1998.\newblock {\rm astro-ph}, {\rm
  9801114}.
\bibitem[{Elmegreen} \& {Lada}<1977>]{EL77}{Elmegreen} B.~G., {Lada} C.~J.,
  1977.\newblock {\rm ApJ}, {\rm 214}, 725.
\bibitem[{Frayer} \& {Brown}<1997>]{FB97}{Frayer} D.~T., {Brown} R.~L.,
  1997.\newblock {\rm ApJSS}, {\rm 113}, 221.
\bibitem[{Heavens} \& {Jimenez}<1999>]{HJ99}{Heavens} A.~F., {Jimenez} R.,
  1999.\newblock {\rm MNRAS}, {\rm }, in press.
\bibitem[{Herbig}<1962>]{H62}{Herbig} G.~H., 1962.\newblock {\rm ApJ}, {\rm
  135}, 965.
\bibitem[{Horner}, {Lada} \& {Lada}<1997>]{HLL97}{Horner} D.~J., {Lada} E.~A.,
  {Lada} C.~J., 1997.\newblock {\rm AJ}, {\rm 113}, 1788.
\bibitem[{Hughes} {\rm et~al.}<1998>]{H+98}{Hughes} D.~H., {Serjeant} S.,
  {Dunlop} J., {Rowan-Robinson} M., {Blain} A., {Mann} R.~G., {Ivison} R.,
  {Peacock} J., {Efstathiou} A., {Gear} W., {Oliver} S., {Lawrence} A.,
  {Longair} M., {Goldschmidt} P., {Jenness} T., 1998.\newblock {\rm Nature},
  {\rm 394}, 241.
\bibitem[{Jimenez} \& {Macdonald}<1996>]{JM96}{Jimenez} R., {Macdonald} J.,
  1996.\newblock {\rm MNRAS}, {\rm 283}, 721.
\bibitem[{Jimenez et al.}<1998>]{JFDTPN99}{Jimenez et al.}, 1998.\newblock {\rm
  astro-ph}, {\rm 9812222}.
\bibitem[{Jimenez} {\rm et~al.}<1998>]{JPMH98}{Jimenez} R., {Padoan} P.,
  {Matteucci} F., {Heavens} A.~F., 1998.\newblock {\rm MNRAS}, {\rm 299}, 123.
\bibitem[Jimenez {\rm et~al.}<1999>]{J+99}Jimenez R., Dunlop J., Padoan P.,
  Peacock J., MacDonald J., Jorgensen U., 1999.\newblock {\rm MNRAS}, {\rm in
  press}.
\bibitem[{Juvela}<1997>]{J97}{Juvela} M., 1997.\newblock {\rm A\&A}, {\rm 322},
  943.
\bibitem[{Kennicutt}<1992>]{K92}{Kennicutt}, Robert~C. J., 1992.\newblock {\rm
  ApJSS}, {\rm 79}, 255.
\bibitem[{Kennicutt}<1998>]{K98}{Kennicutt} R.~C., 1998.\newblock {\rm ApJ},
  {\rm 498}, 541.
\bibitem[{Kurucz}<1992>]{Kurucz_92}{Kurucz} R., 1992.\newblock In: {\it
  CDROM13}.
\bibitem[{Lada} \& {Gautier}<1982>]{LG82}{Lada} C.~J., {Gautier}, T.~N. I.,
  1982.\newblock {\rm ApJ}, {\rm 261}, 161.
\bibitem[{Lada} \& {Lada}<1995>]{LL95}{Lada} E.~A., {Lada} C.~J.,
  1995.\newblock {\rm AJ}, {\rm 109}, 1682.
\bibitem[{Lada}, {Alves} \& {Lada}<1996>]{LAL96}{Lada} C.~J., {Alves} J.,
  {Lada} E.~A., 1996.\newblock {\rm AJ}, {\rm 111}, 1964.
\bibitem[{Lada} {\rm et~al.}<1991>]{LEDG91}{Lada} E.~A., {Evans}, Neal~J. I.,
  {Depoy} D.~L., {Gatley} I., 1991.\newblock {\rm ApJ}, {\rm 371}, 171.
\bibitem[{Lada} {\rm et~al.}<1994>]{L+94}{Lada} C.~J., {Lada} E.~A., {Clemens}
  D.~P., {Bally} J., 1994.\newblock {\rm ApJ}, {\rm 429}, 694.
\bibitem[{Lada}, {Lada} \& {Myers}<1993>]{LLM93}{Lada} E.~F., {Lada} E.~A.,
  {Myers} P.~C., 1993.\newblock {\rm ApJ}, {\rm 410}, 168.
\bibitem[{Lada}, {Young} \& {Greene}<1993>]{LYG93}{Lada} C.~J., {Young} E.~T.,
  {Greene} T.~P., 1993.\newblock {\rm ApJ}, {\rm 408}, 471.
\bibitem[{Leitherer et al.}<1996>]{L+96}{Leitherer et al.}, 1996.\newblock {\rm
  PASP}, {\rm 108}, 996.
\bibitem[{Lowenthal} {\rm et~al.}<1997>]{Lo+97}{Lowenthal} J.~D., {Koo} D.~C.,
  {Guzman} R., {Gallego} J., {Phillips} A.~C., {Faber} S.~M., {Vogt} N.~P.,
  {Illingworth} G.~D., {Gronwall} C., 1997.\newblock {\rm ApJ}, {\rm 481}, 673.
\bibitem[{Madau}<1997>]{M97}{Madau} P., 1997.\newblock {\rm astro-ph}, {\rm
  9709147}.
\bibitem[{Maiz-Apellaniz et al.}<1998>]{MA98}{Maiz-Apellaniz et al.},
  1998.\newblock {\rm astro-ph}, {\rm 9812138}.
\bibitem[{Mathis}, {Rumpl} \& {Nordsieck}<1977>]{MRN77}{Mathis} J.~S., {Rumpl}
  W., {Nordsieck} K.~H., 1977.\newblock {\rm ApJ}, {\rm 217}, 425.
\bibitem[{Matteucci}, {Ponzone} \& {Gibson}<1998>]{MPG98}{Matteucci} F.,
  {Ponzone} R., {Gibson} B.~K., 1998.\newblock {\rm A\&A}, {\rm 335}, 855.
\bibitem[{Mayya} \& {Prabhu}<1996>]{MP96}{Mayya} Y.~D., {Prabhu} T.~P.,
  1996.\newblock {\rm AJ}, {\rm 111}, 1252.
\bibitem[{Meurer}, {Heckman} \& {Calzetti}<1999>]{MHC99}{Meurer} G.~R.,
  {Heckman} T.~M., {Calzetti} D., 1999.\newblock {\rm ApJ}, {\rm 521}, 64.
\bibitem[{Meyer} {\rm et~al.}<1994>]{MJHC94}{Meyer} D.~M., {Jura} M., {Hawkins}
  I., {Cardelli} J.~A., 1994.\newblock {\rm ApJL}, {\rm 437}, L59.
\bibitem[{Padoan} \& {Nordlund}<1999>]{PN98}{Padoan} P., {Nordlund} A.,
  1999.\newblock {\rm ApJ}, {\rm submitted}.
\bibitem[Padoan {\rm et~al.}<1998>]{PJBN98}Padoan P., Juvela M., Bally J.,
  Nordlund {\AA}., 1998.\newblock {\rm ApJ}, {\rm 504}, 300.
\bibitem[{Padoan} {\rm et~al.}<1999>]{PBBJN99}{Padoan} P., {Bally} J.,
  {Billawala} Y., {Juvela} M., {Nordlund} A., 1999.\newblock {\rm ApJ}, {\rm
  submitted}.
\bibitem[Padoan, Jimenez \& Jones<1997>]{PJJ97}Padoan P., Jimenez R., Jones B.
  J.~T., 1997.\newblock {\rm MNRAS}, {\rm 285}, 711.
\bibitem[Padoan, Nordlund \& Jones<1997>]{PNJ97}Padoan P., Nordlund {\AA}.,
  Jones B., 1997.\newblock {\rm ApJ}, {\rm 474}, 730.
\bibitem[{Peacock et al.}<1999>]{Peacock+99}{Peacock et al.}, 1999.\newblock
  {\rm MNRAS}, {\rm submitted}.
\bibitem[{Pettini et al.}<1999>]{PSADG99}{Pettini et al.}, 1999.\newblock {\rm
  astro-ph}, {\rm 9908007}.
\bibitem[{Pettini} {\rm et~al.}<1998>]{PKSDAG98}{Pettini} M., {Kellogg} M.,
  {Steidel} C.~C., {Dickinson} M., {Adelberger} K.~L., {Giavalisco} M.,
  1998.\newblock {\rm ApJ}, {\rm 508}, 539.
\bibitem[{Spinrad} {\rm et~al.}<1997>]{Spinrad+97}{Spinrad} H., {Dey} A.,
  {Stern} D., {Dunlop} J., {Peacock} J., {Jimenez} R., {Windhorst} R.,
  1997.\newblock {\rm ApJ}, {\rm 484}, 581.
\bibitem[{Steidel et al.}<1998>]{S+98}{Steidel et al.}, 1998.\newblock {\rm
  astro-ph}, {\rm 9811399}.
\bibitem[{Steidel}, {Pettini} \& {Hamilton}<1995>]{SPH95}{Steidel} C.~C.,
  {Pettini} M., {Hamilton} D., 1995.\newblock {\rm AJ}, {\rm 110}, 2519.
\bibitem[{Testor} \& {Niemela}<1998>]{TN98}{Testor} G., {Niemela} V.,
  1998.\newblock {\rm A\&ASS}, {\rm 130}, 527.
\bibitem[{Whitworth}<1979>]{W79}{Whitworth} A., 1979.\newblock {\rm MNRAS},
  {\rm 186}, 59.
\bibitem[{Wilcots}<1994>]{W94}{Wilcots} E.~M., 1994.\newblock {\rm AJ}, {\rm
  107}, 1338.
\bibitem[{Williams} {\rm et~al.}<1996>]{W+96}{Williams} R.~E., {Blacker} B.,
  {Dickinson} M., {Dixon} W. V.~D., {Ferguson} H.~C., {Fruchter} A.~S.,
  {Giavalisco} M., {Gilliland} R.~L., {Heyer} I., {Katsanis} R., {Levay} Z.,
  {Lucas} R.~A., {McElroy} D.~B., {Petro} L., {Postman} M., {Adorf} H.-M.,
  {Hook} R., 1996.\newblock {\rm AJ}, {\rm 112}, 1335.
\bibitem[{Xu} \& {De Zotti}<1989>]{XZ89}{Xu} C., {De Zotti} G., 1989.\newblock
  {\rm A\&A}, {\rm 225}, 12.
\bibitem[{Yoshii} \& {Peterson}<1994>]{YP94}{Yoshii} Y., {Peterson} B.~A.,
  1994.\newblock {\rm ApJ}, {\rm 436}, 551.
\bibitem[{Young} \& {Scoville}<1991>]{YS91}{Young} J.~S., {Scoville} N.~Z.,
  1991.\newblock {\rm ARA\&A}, {\rm 29}, 581.
\bibitem[{Zepf}<1997>]{Z97}{Zepf} S.~E., 1997.\newblock {\rm Nature}, {\rm
  390}, 377.
\end{thebibliography}

\clearpage
\noindent{\bf Figure captions}

\noindent{\bf Figure 1:} Three-dimensional visualizations of the density field
in highly super-sonic magneto-hydrodynamic turbulent flows, computed in a
128$^3$ grid. The left panel shows a flow with super-alfv\'{e}nic velocity
dispersion (kinetic energy in excess of the magnetic energy), while the right
panel shows an equipartition flow (kinetic energy of the order of the magnetic
energy). High-density filaments are partially aligned with the magnetic field
in the equipartition case, because high-density structures are stretched by
motions along magnetic field-lines.

\noindent{\bf Figure 2:} The time evolution of several chemical species for a
starburst of duration 0.4 Gyr. Note that at all times O/C$>$1.

\noindent{\bf Figure 3:} Time evolution of the Spectral energy distribution
(SED) for a starburst of duration 0.3 Gyr, corresponding to a mass of 1
$\times$ 10$^{11}$ M$_{\odot}$. The last panel (age=0.3 Gyr), show the SED without
the dust emission. The $^{12}$CO molecular lines are also plotted (see
text). Reddening has been computed using the method developed in \S 3.2.

\noindent{\bf Figure 4:} Sub--mm SED of the two high--redshift radio--galaxies
4C41.17 and 8C1435 (diamonds), from Archibald (1999, private comm. PhD Thesis,
University of Edinburgh). The continuous line is the SED of the starburst
model, appropriately redshifted, and computed for the age of $10^7$~yr, for
the fit to 8C1435, and $10^8$~yr, for the fit to 4C41.17.

\noindent{\bf Figure 5:} Time evolution of the luminosity of the $^{12}$CO
lines.

\noindent{\bf Figure 6:} Four top panels: SED of simple stellar populations
(see text) of three different ages, 10, 20 and 30 Myr. Each panel corresponds
to a different value of the metallicity. Two bottom panels: the same as above,
but only for two values of age and two values of metallicity, and with spectra
normalized at 1800 {\AA}.

\noindent{\bf Figure 7:} The observed spectrum of the Lyman-limit galaxy
(1512-cB58) (thin continuous line), compared with our starburst model (thick
continuous line). The dashed line shows the SED of the model without
extinction, and the dotted line the model with extinction, but with a delay of
30 Myr for the stars to become visible.  Since most of the absorption features
seen in the spectrum of 1512-cB58 are due to the interstellar medium that is
not included in our theoretical model (it includes only stellar absorption
features), we choose to plot the SED of 1512-cB58 at higher resolution, to
emphasize the fact that we are only attempting to fit the continuum. In the
case of super--solar metallicity (upper panel), the absorption features due to
the stars alone have equivalent widths bigger than those observed in
1512-cB58, which are primarily due to interstellar medium. The metallicities
of Lyman-limit objects is therefore unlikely to be larger than solar.

\noindent{\bf Figure 8:} $L_{FIR}/L_{UV}$ ratio versus the spectral
slope $\alpha$, in the starburst model (continuous thick line), and in 
a screen model (dashed line). The value of the spectral slope $\alpha$
is not computed in our model for an age smaller than 15 Myr (see text),
that corresponds to $\alpha\approx 0$.

\noindent{\bf Figure 9:} Comparison of our starburst model with the data for
nearby starburst from \scite{C97} (diamonds). The range of IR
colors and $F(0.16 \mu m)/F(2.18 \mu m)$ can be reproduced by our model for
the measured range of metallicities.

\noindent{\bf Figure 10:} Upper panel: Best fit to the SED of the 
$z \approx 1.5$ radio-galaxy 53W069. Lower panel: The SED of the model
is split into the component due to stars with metallicity $Z>Z_{\odot}$
(dotted line), and the component due to stars with sub--solar metallicity,
$Z<Z_{\odot}$. The low metallicity stars, born at the beginning of the 
starburst, give the largest contribution to the FUV light, but would not
fit alone the SED at longer wavelengths.

\noindent{\bf Figure 11:} Redshift evolution for selected photometric
colors. Due to the extended period of star formation (0.6 Gyr in this case),
colors are never as red as in models with instantaneous star formation. Note
the excellent fit to the reddest known ellipticals at $z \sim 1.5$, and the
fact that the reddest colors occur at moderate redshifts, in contrast with
models with instantaneous star formation.

\noindent{\bf Figure 12:} SED of the starburst model after 13 Gyr, 
compared with the SED of E/S0 and Sa galaxies from the Kennicutt catalog
\cite{K92}.

\clearpage
\begin{table*}
\begin{center}
\caption{Main properties of Galactic Dark Clouds.}
\begin{tabular}{lllll}
\hline
Name & IR excess & Age/(10$^6$yr) & A$_V$/mag & main reference \\
\hline\hline 
CL1 (M17) & 100\% & 1      & 8.0      & \scite{LEDG91} \\
NGC 133   & 61\%  & 1-2    & 7.0      & \scite{LAL96} \\
NGC 2264  & 50\%  & 5      & 4.5      & \scite{LYG93} \\
IC 348    & 21\%  & 5-7    & 4.5      & \scite{LL95} \\
NGC 2282  & 9\%   & 15     & 1.4      & \scite{HLL97} \\
\hline 
\end{tabular}
\end{center}
\end{table*}

\clearpage
\begin{figure}
\centerline{
\psfig{figure=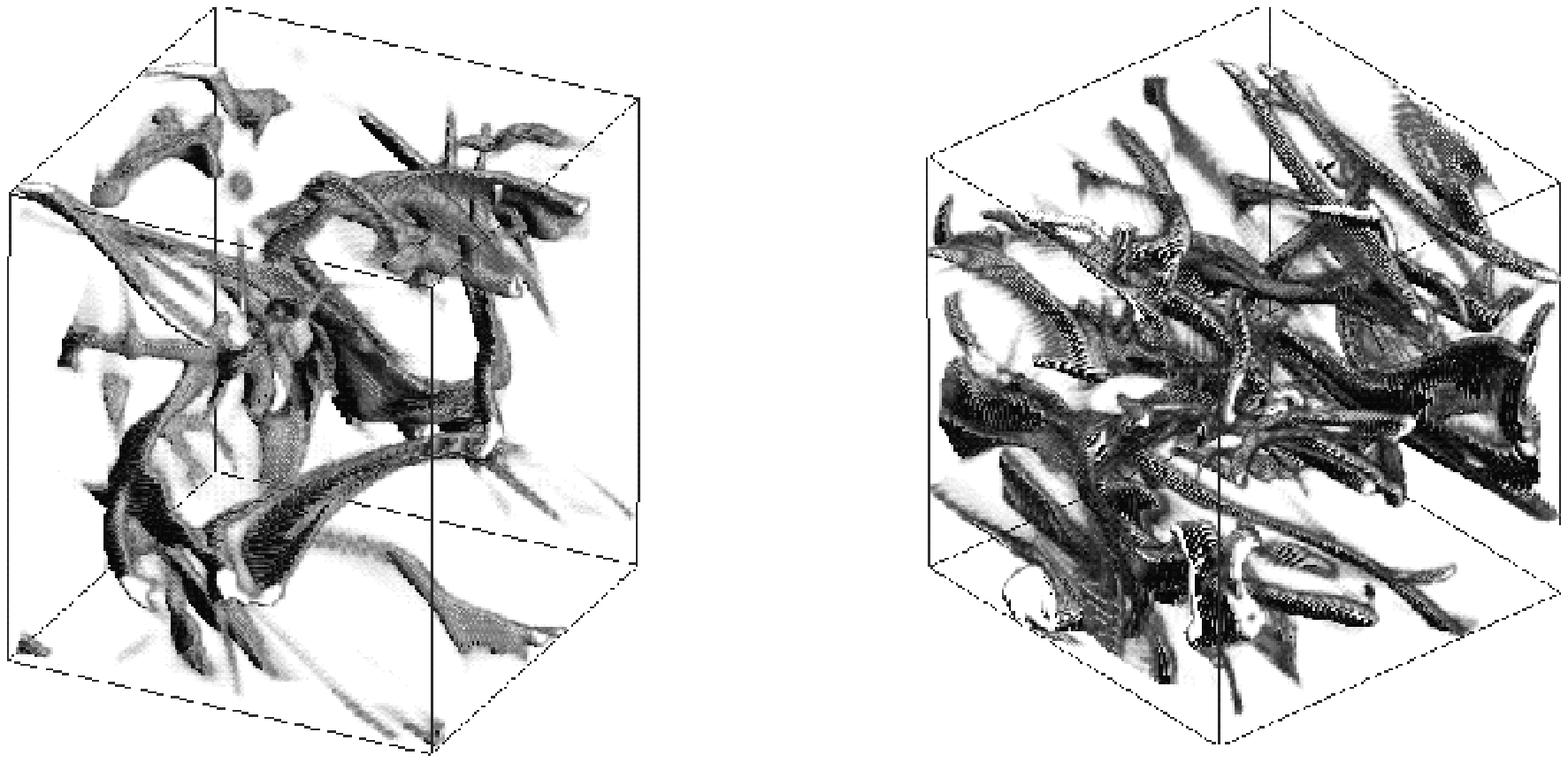,height=9cm,angle=0}}
\caption{}
\end{figure}

\clearpage
\begin{figure}
\centerline{
\psfig{figure=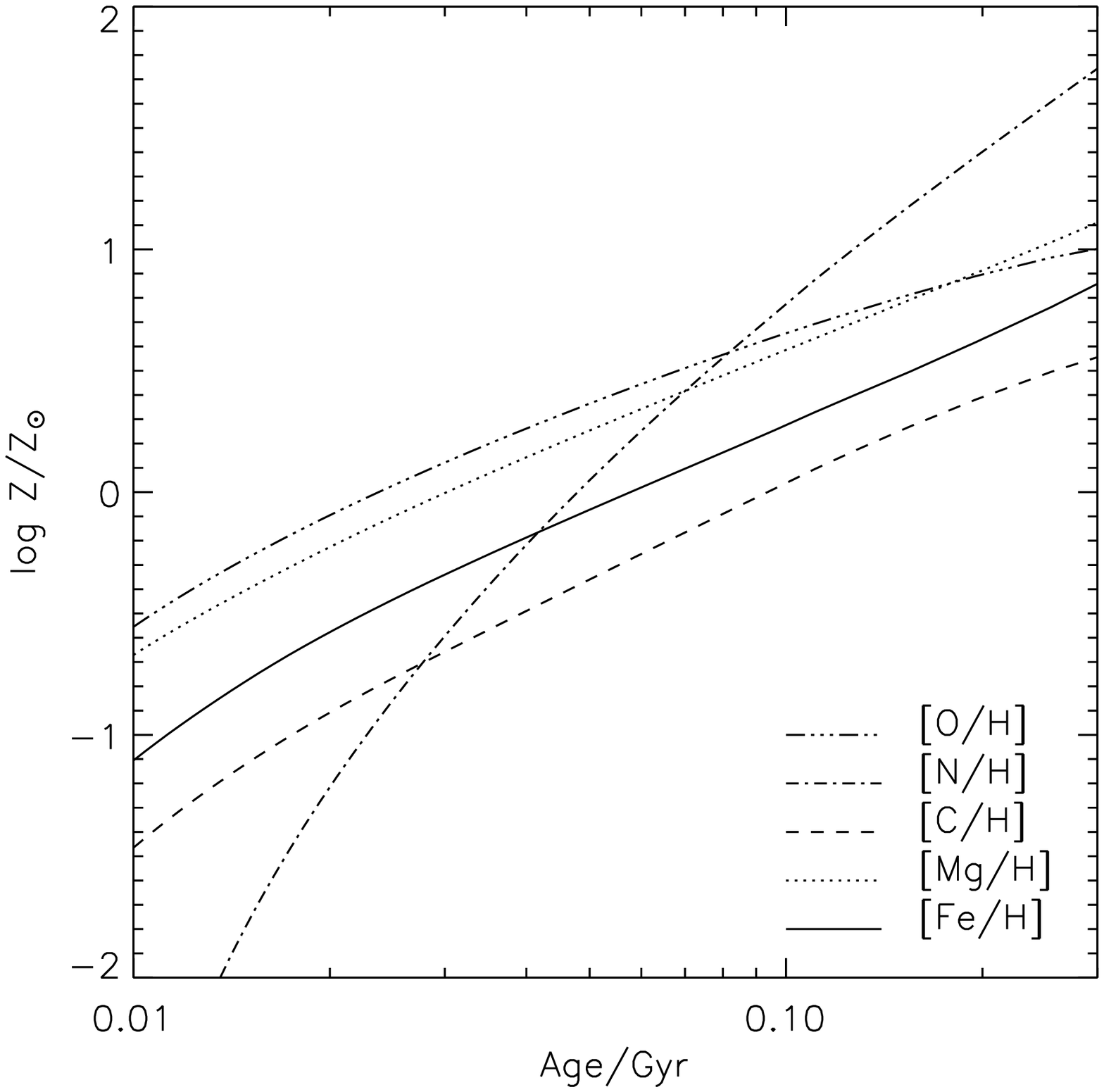,height=12cm,angle=0}}
\caption{}
\end{figure}

\clearpage
\begin{figure}
\centerline{
\psfig{figure=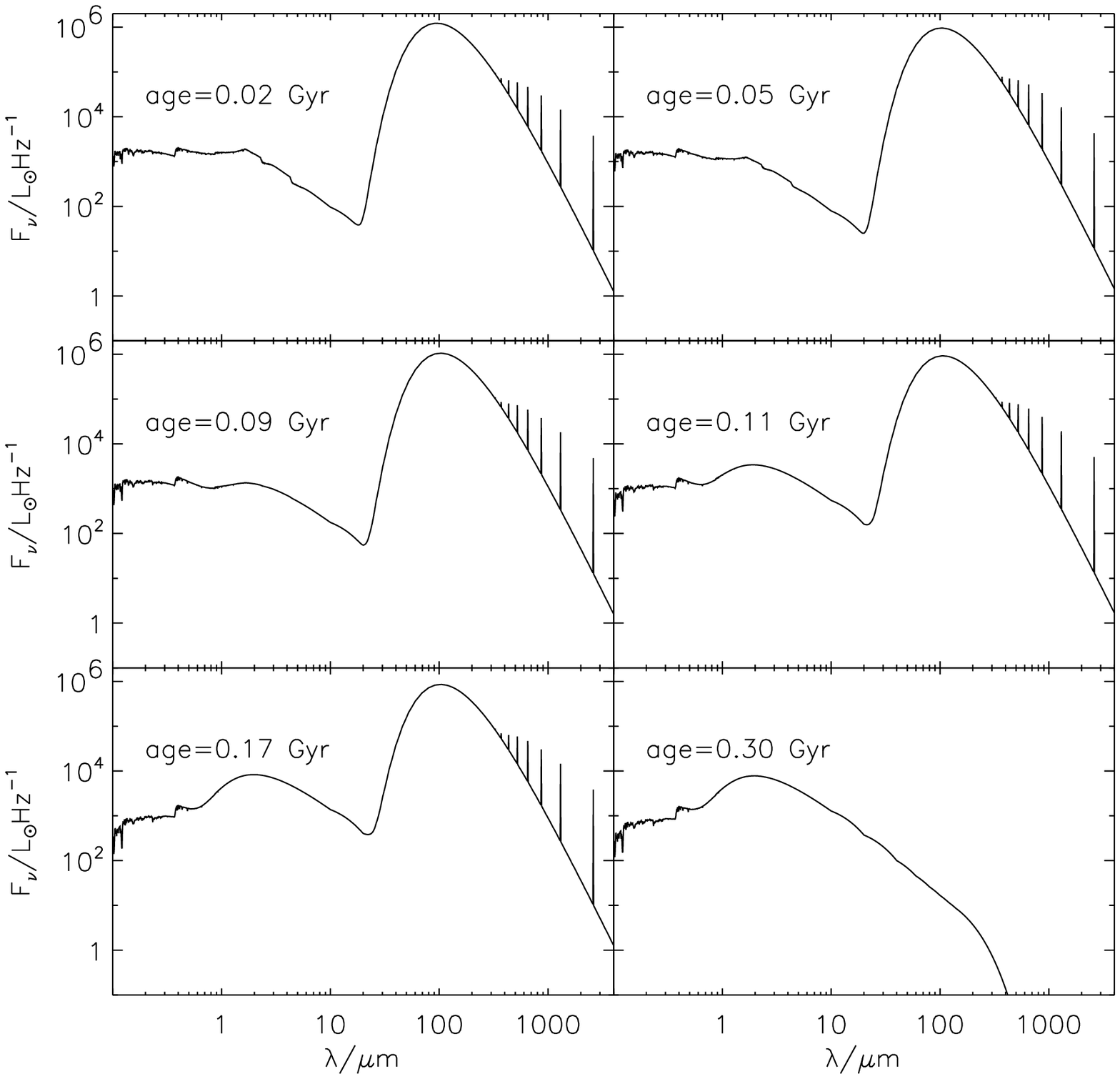,height=18cm,angle=0}}
\caption{}
\end{figure}

\clearpage
\begin{figure}
\centerline{
\psfig{figure=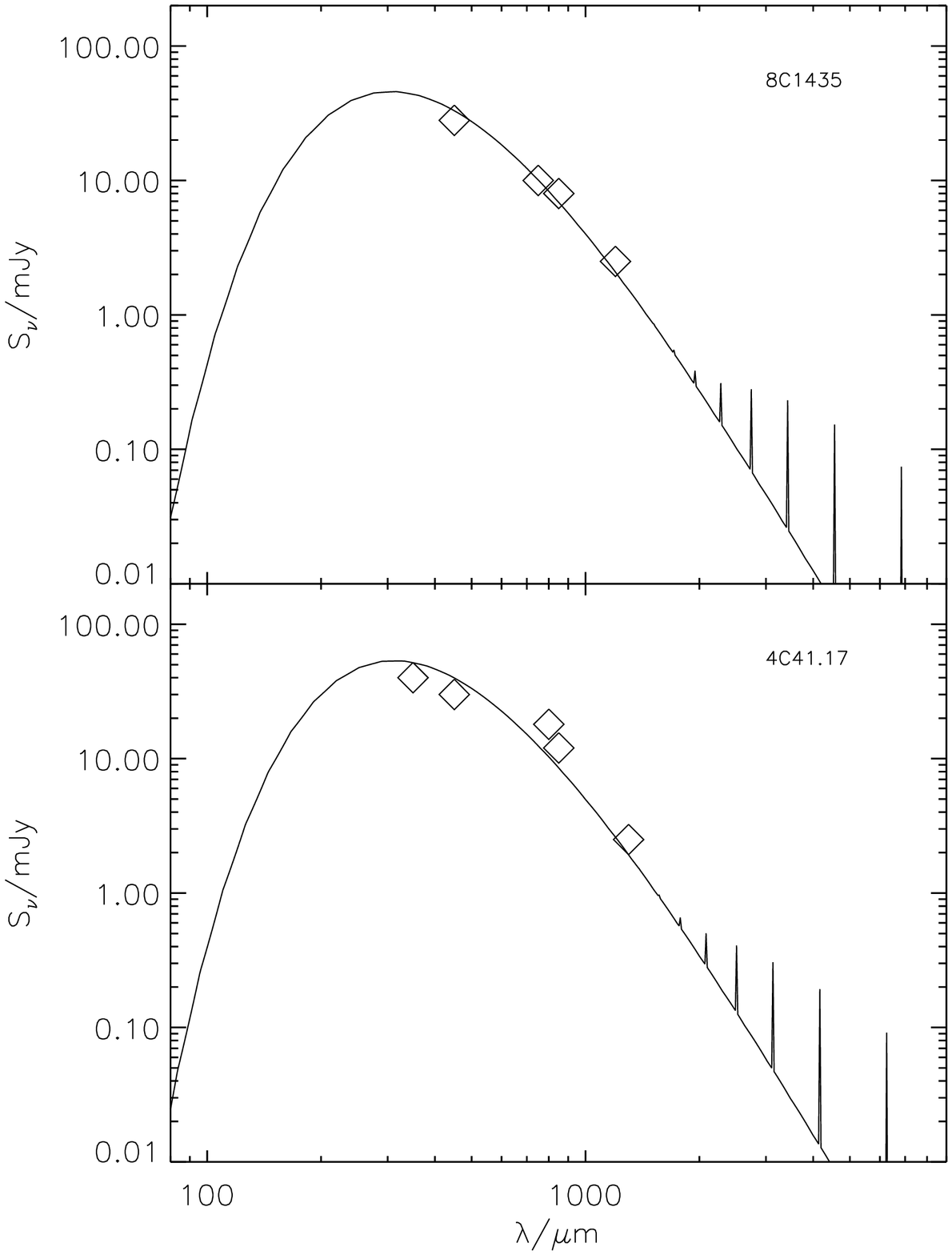,height=18cm,angle=0}}
\caption{}
\end{figure}

\clearpage
\begin{figure}
\centerline{
\psfig{figure=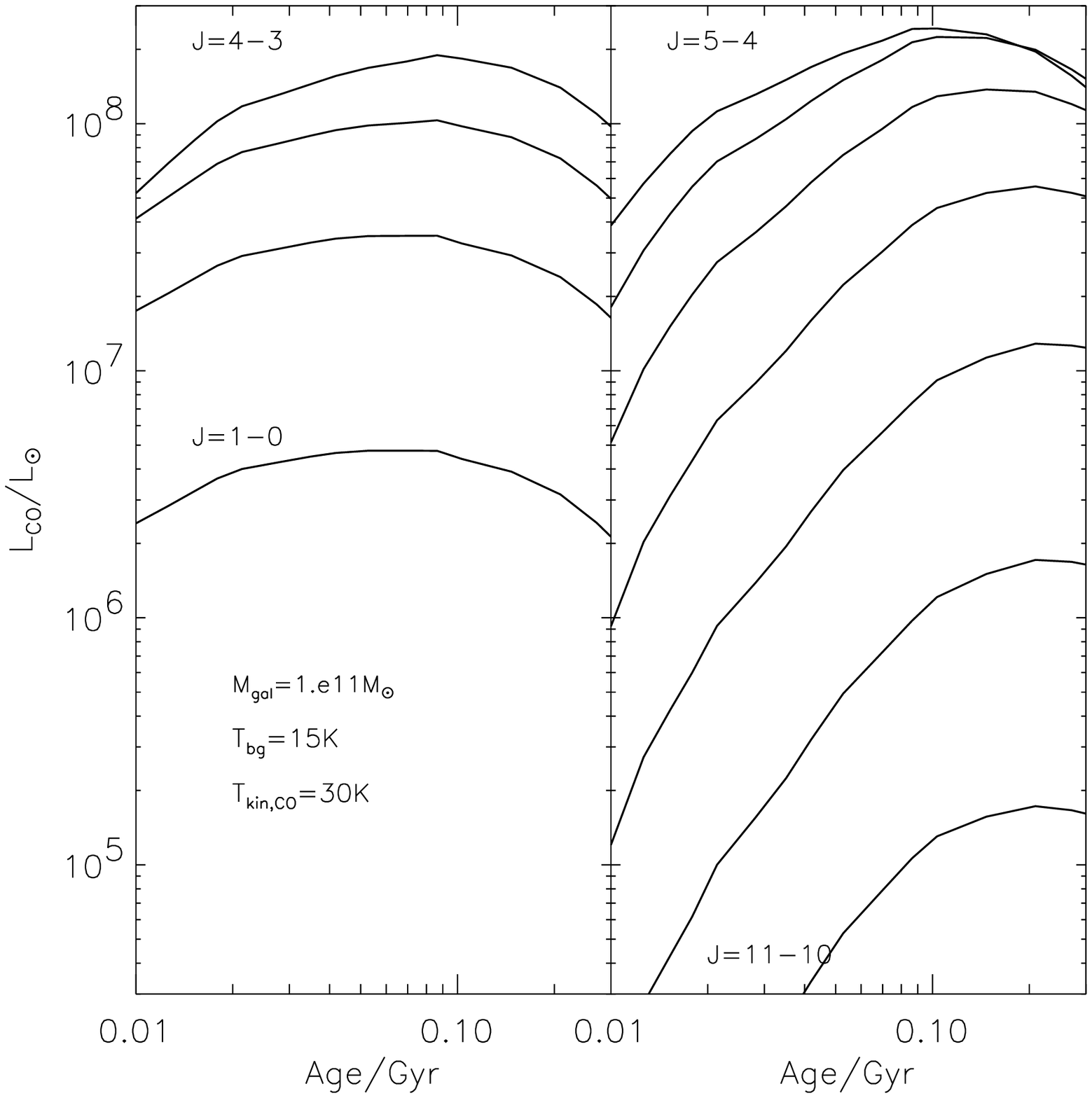,height=12cm,angle=0}}
\caption{}
\end{figure}

\clearpage
\begin{figure}
\centerline{
\psfig{figure=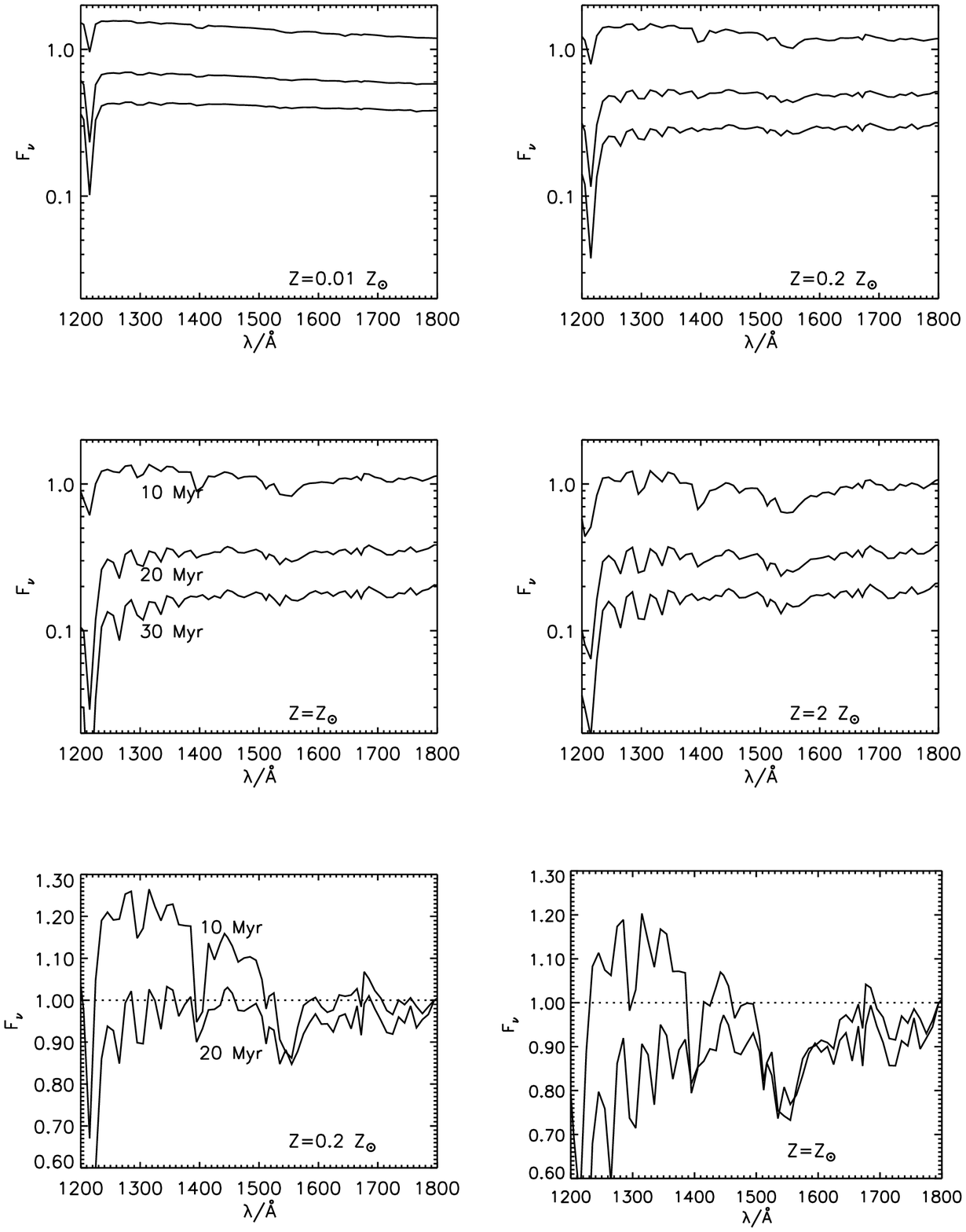,height=18cm,angle=0}}
\caption{}
\end{figure}

\clearpage
\begin{figure}
\centerline{
\psfig{figure=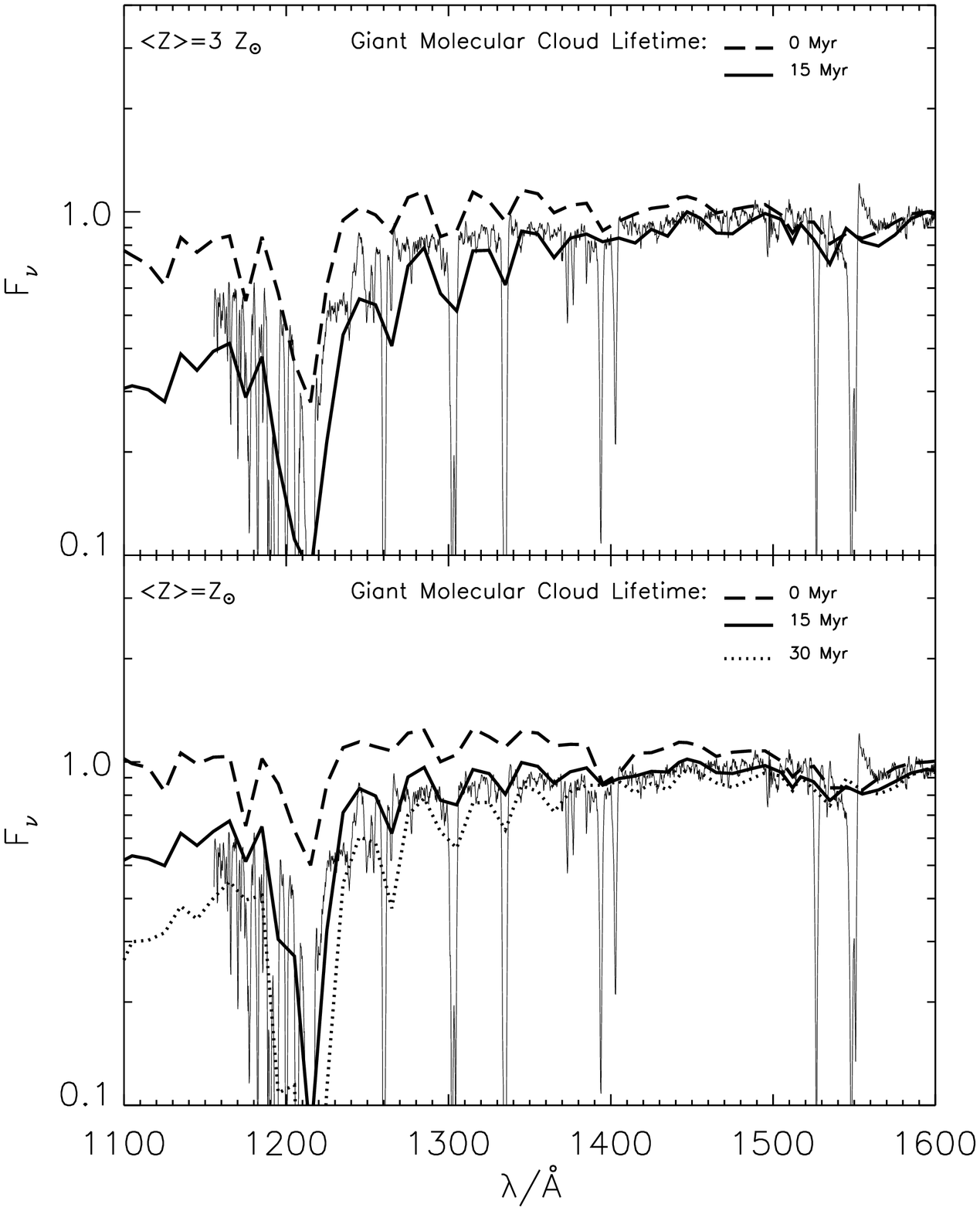,height=18cm,angle=0}}
\caption{}
\end{figure}

\clearpage
\begin{figure}
\centerline{
\psfig{figure=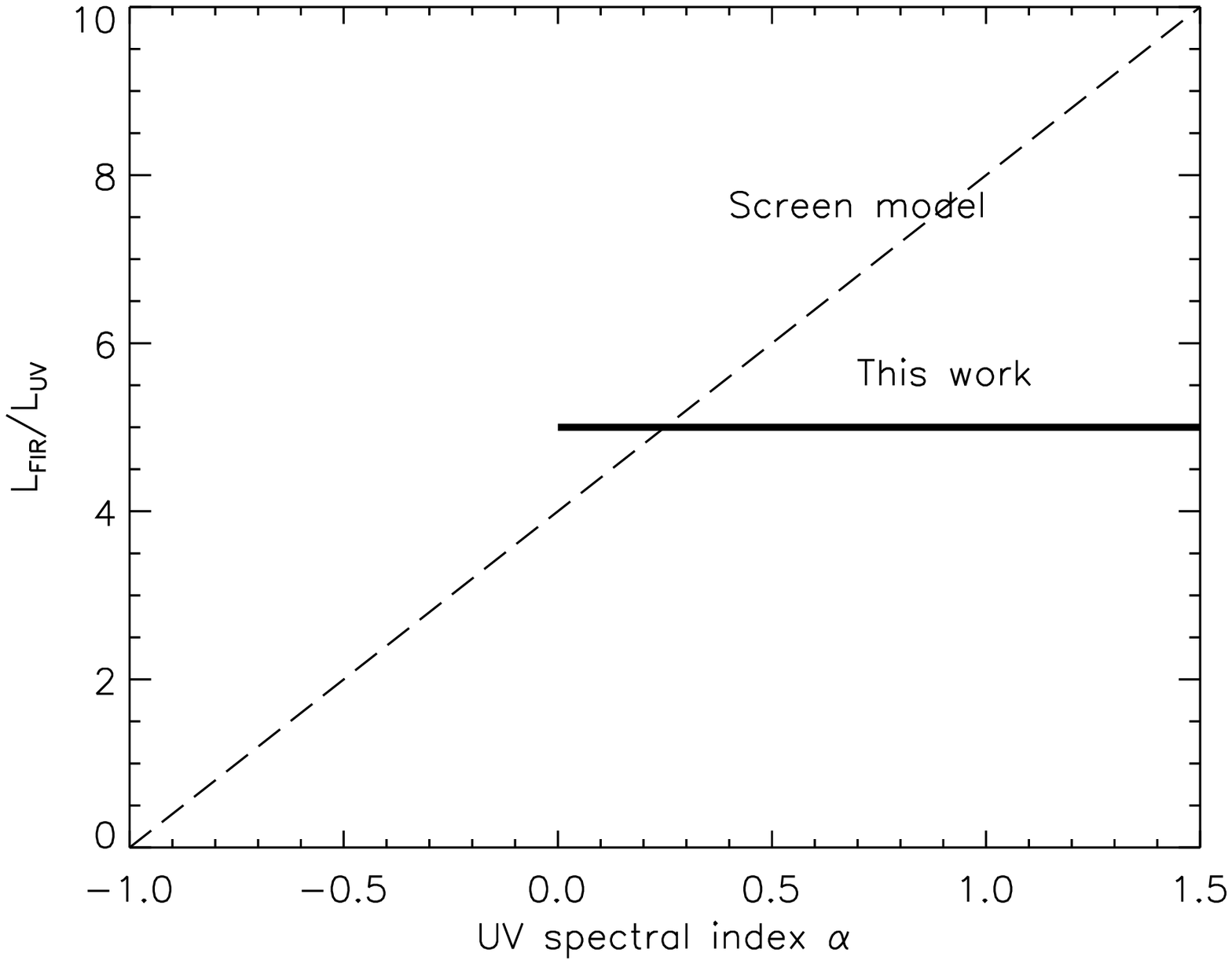,height=14cm,angle=0}}
\caption{}
\end{figure}

\clearpage
\begin{figure}
\centerline{
\psfig{figure=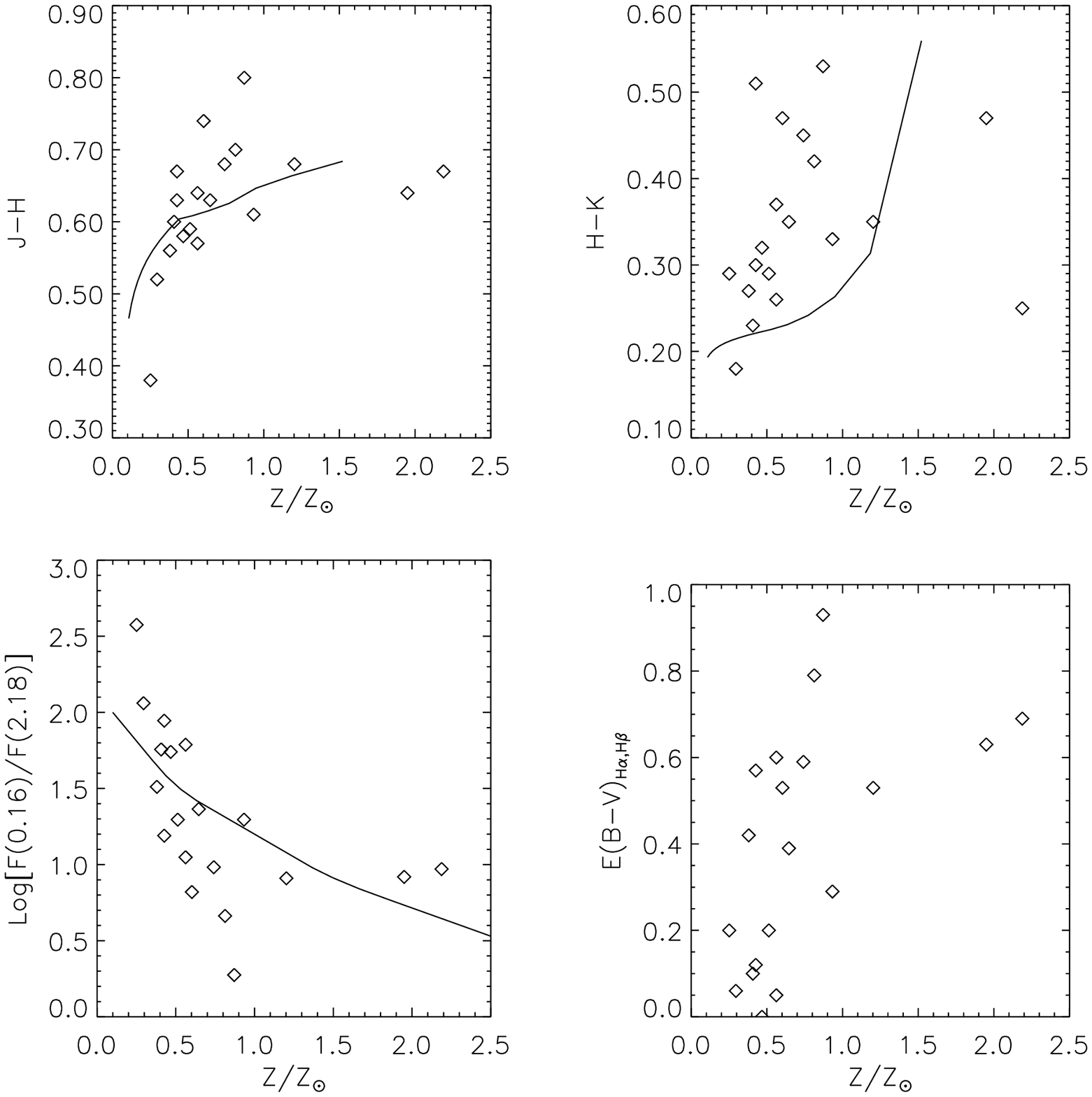,height=18cm,angle=0}}
\caption{}
\end{figure}

\clearpage
\begin{figure}
\centerline{
\psfig{figure=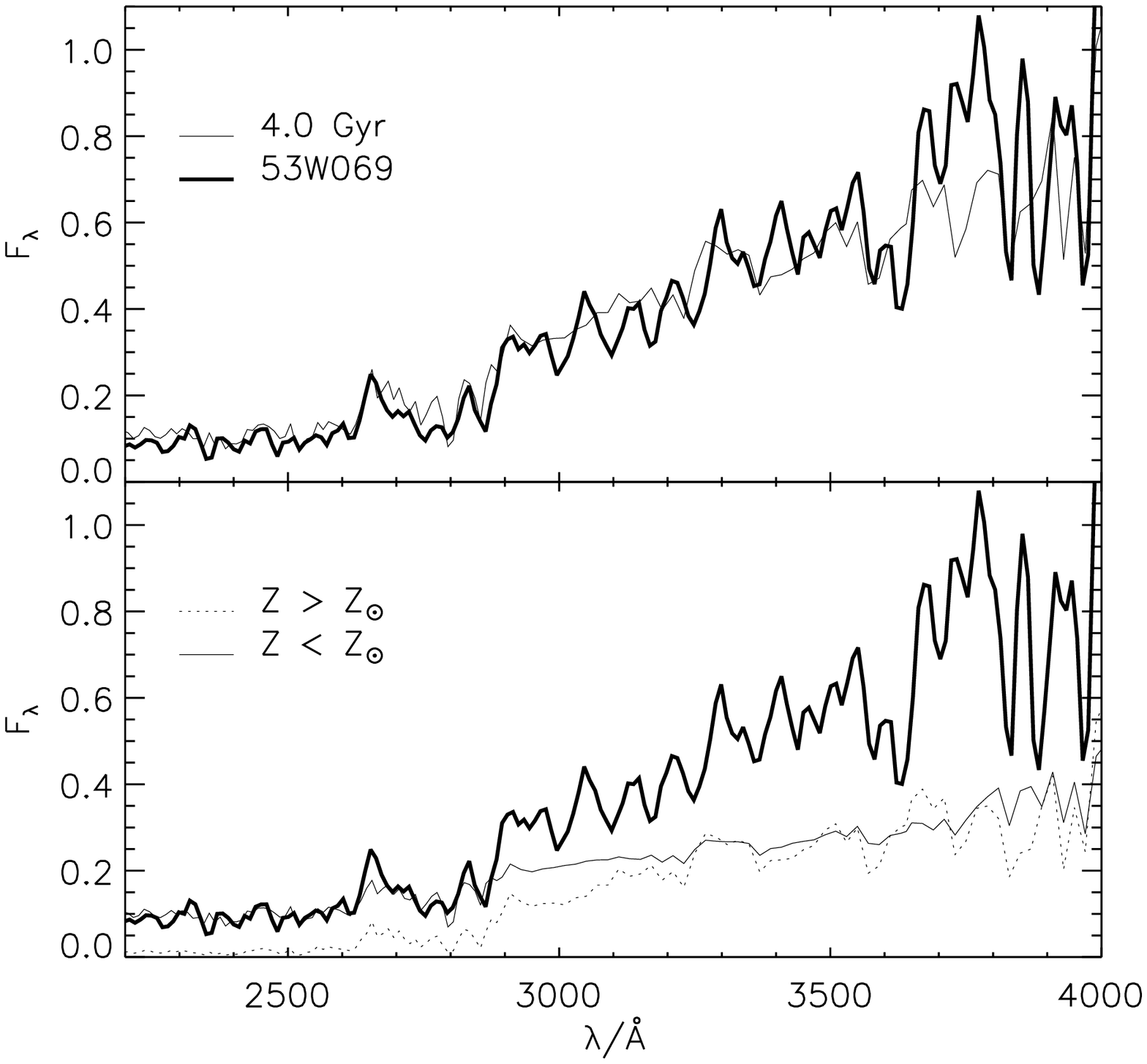,height=14cm,angle=0}}
\caption{}
\end{figure}

\clearpage
\begin{figure}
\centerline{
\psfig{figure=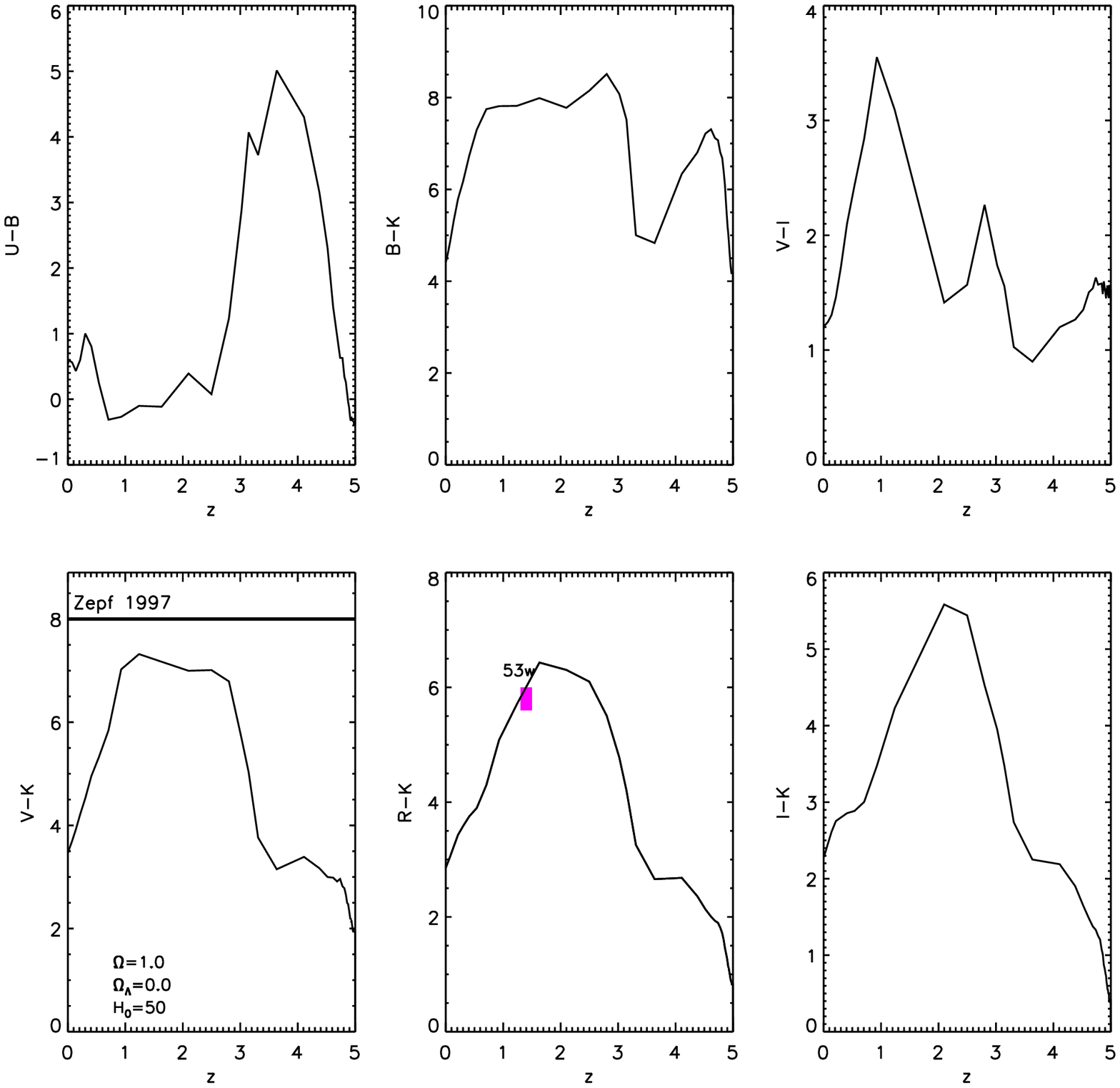,height=18cm,angle=0}}
\caption{}
\end{figure}     

\clearpage
\begin{figure}
\centerline{
\psfig{figure=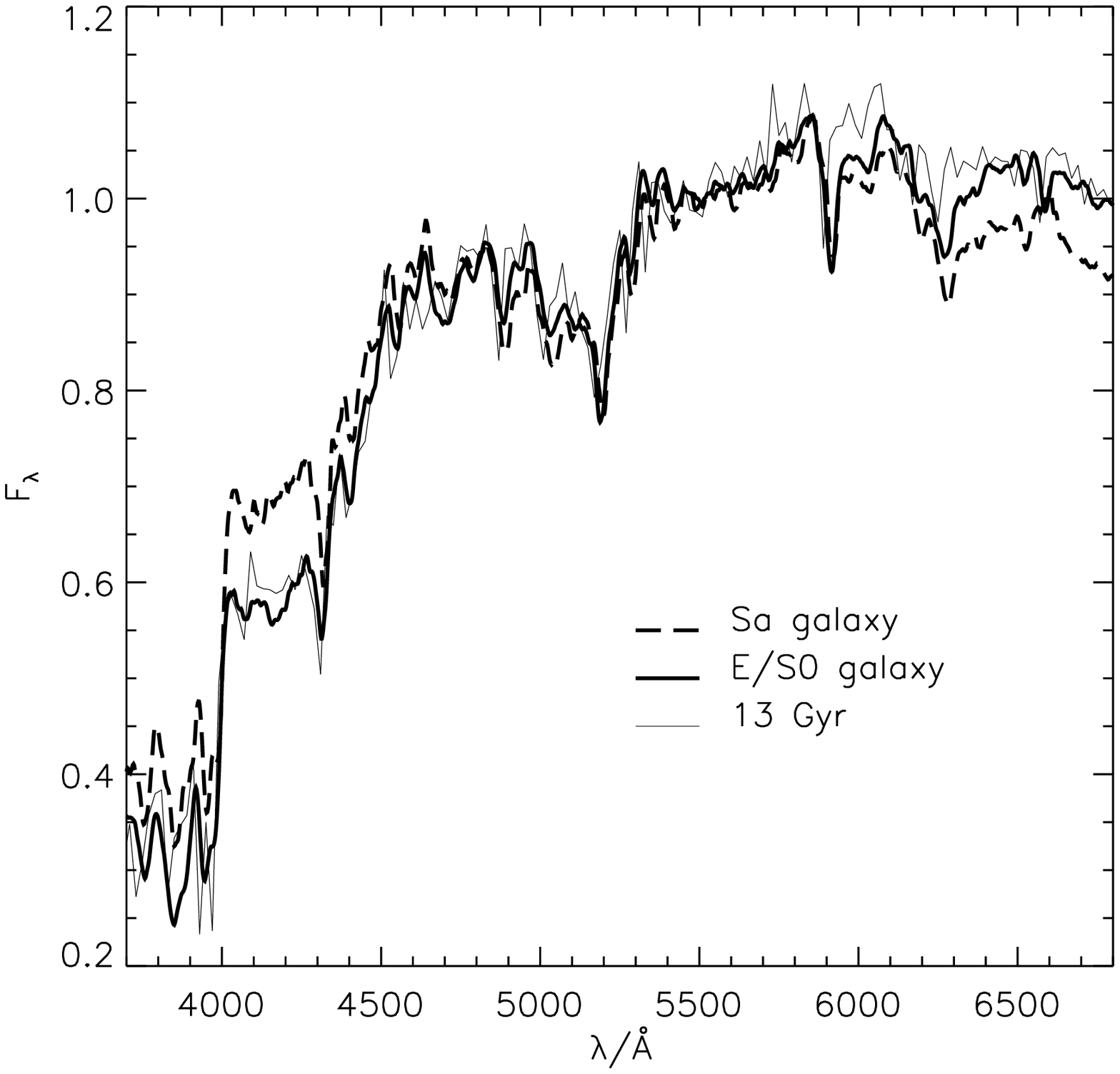,height=12cm,angle=0}}
\caption{}
\end{figure}

\end{document}